\documentclass[journal]{IEEEtran}

\usepackage{amsmath,amsfonts}
\usepackage{algorithmic}
\usepackage{algorithm}
\usepackage{array}
\usepackage[caption=false,font=normalsize,labelfont=sf,textfont=sf]{subfig}
\usepackage{textcomp}
\usepackage{stfloats}
\usepackage{url}
\usepackage{verbatim}
\usepackage{graphicx}
\usepackage{cite}
\usepackage{graphicx}
\usepackage{float}
\usepackage{algorithm}
\usepackage{multirow}
\usepackage{colortbl} 
\usepackage{soul,color}
\usepackage{hyperref}
\usepackage{tikz}
\usepackage{amsmath}
\usepackage[framemethod=TikZ]{mdframed}
\usepackage{xspace}

\usepackage{minibox}
\newcounter{takeaway}
\newcommand*{\takeaway}[1]{%
    \stepcounter{takeaway}%
    \begin{center}
    \vspace{-4pt}
    \minibox[frame, rule=1pt,pad=3pt]{
        \begin{minipage}[t]{0.95\columnwidth}
        \textbf{Takeaway~\arabic{takeaway}:} \textit{#1}
        \end{minipage}
    }
    \vspace{-4pt}
    \end{center}
}

\begin{document}
%

\title{Repurposing and Evaluating the (In)Feasibility of Dataset Poisoning enabled Watermarking for Contrastive Learning}

\author{Zhiyang~Dai,
        Yansong~Gao,~\IEEEmembership{Senior Member,~IEEE,}
        Boyu~Kuang,
        Haodong~Li,
        Qi~Chang,
        Gaurav~Varshney, \\
    Derek~Abbott,~\IEEEmembership{Life Fellow,~IEEE}
        \textit{and} Anmin~Fu,~\IEEEmembership{Member,~IEEE}
 
\thanks{This work is supported by National Natural Science Foundation of China (62372236, 62402223), Funded by Open Foundation of Shenzhen Key Laboratory of Media Security (SYSPG20241211174032004) and partially supported by Australian Research Council FL240100217. \emph{(Zhiyang Dai and Yansong Gao are co-first authors.) (Corresponding author: Anmin Fu.)}

Z.~Dai and B.~Kuang are with the School of Cyber Science and Engineering, Nanjing University of Science and Technology, Nanjing 210094, China. E-mail: \{dzy; kuang\}@njust.edu.cn

Y.~Gao is with the University of Western Australia, Perth, Australia. E-mail: garrison.gao@uwa.edu.au

H.~Li and Q.~Chang are with the College of Electronics and Information Engineering, Shenzhen University, Shenzhen, China. E-mail: \{lihaodong; qichang\}@szu.edu.cn

G.~Varshney is currently with the Department of Computer Science and Engineering, Indian Institute of Technology Jammu, Jammu, India.  E-mail: gaurav.varshney@iitjammu.ac.in

D. Abbott is with the School of Electrical and Mechanical Engineering, Adelaide University, Adelaide, SA 5005, Australia. E-mail: derek.abbott@adelaide.edu.au.

A.~Fu is with the School of Computer Science and Engineering, Nanjing University of Science and Technology, Nanjing 210094, China. E-mail: fuam@njust.edu.cn
}}

\markboth{Journal of \LaTeX\ Class Files,~Vol.~X, No.~X, May~2026}%
{Shell \MakeLowercase{\textit{et al.}}: A Sample Article Using IEEEtran.cls for IEEE Journals}

\maketitle

\begin{abstract}
Contrastive learning (CL) is a core category of self-supervised learning (SSL) that alleviates the dependence on costly human annotations by leveraging supervisory signals automatically derived from data. 
In practice, constructing large-scale in-house CL datasets is often infeasible, necessitating reliance on third-party sources or internet-scale data collection. Meanwhile, recent studies have demonstrated that CL models are vulnerable to backdoor attacks via dataset poisoning, raising significant security concerns. However, the generalization and robustness of such attacks remain underexplored, as inducing effective backdoors in CL is inherently more challenging than in supervised learning. Motivated by this gap, we first conduct a systematic evaluation of existing data-poisoning-only backdoor attacks against CL within a unified experimental framework. Our analysis reveals noteworthy limitations, including poor dataset adaptability, low attack success rates, limited architecture portability, and restrictive attack assumptions (e.g., requiring knowledge of downstream tasks).

Interestingly, we observe that trigger-carrying samples exhibit distinguishable \textit{statistical divergence} from clean sample counterparts. This observation inspires us to explore the dual use of such divergence as a watermark signal for protecting dataset intellectual property (IP). However, directly repurposing this weak signal is non-trivial due to the inherently low attack success rates, which are overcome by statistical verification using a unified density metric instead of relying on the attack success rate. In addition, to satisfy different types of CL foundation model usage, we propose a multi-level watermarking scheme that enables IP verification to flexibly adapt to different levels of output information accessible in CL: feature-level representations (from encoders), soft-label outputs (from downstream classifiers), and hard-label predictions (from downstream classifiers). Extensive experiments demonstrate that some data poisoning backdoor attacks can be repurposed as effective watermarking methods, exhibiting distinct trade-offs across fidelity, verifiability, and robustness. Importantly, this work shows that weak backdoor effects can be transformed into reliable signals for dataset IP protection in challenging CL settings.

\end{abstract}


\IEEEpeerreviewmaketitle

\section{Introduction}
\IEEEPARstart{S}{elf-Supervised} Learning (SSL) can automatically capture supervisory signals directly from massive unlabeled data, learning intrinsic features and patterns, thereby avoiding the tedious and costly manual annotations required in traditional supervised learning. In particular, Contrastive Learning (CL), a core learning paradigm of SSL \cite{survey2024, ssl}, has powered numerous landmark models such as DINOv2 \cite{DINOv2} and CLIP \cite{CLIP}, significantly advancing the development of general artificial intelligence. However, the core of CL is built upon a \it big data \rm foundation, where its performance and generalization capability rely heavily on the scale, quality, and diversity of the training data \cite{Data-Efficient, When}. This inevitably demands large-scale datasets, for which in-house creation is often impractical.

To satisfy this demand, both industry and academia primarily rely on web crawling \cite{cdi} to collect content, including text, images from public platforms, digital libraries, and various databases \cite{OBELICS}. While this approach is efficient and cost-effective, it is vulnerable to data poisoning attacks. Malicious parties can inject poisoned samples into publicly available datasets, thereby enabling backdoor attacks against CL models trained on them. More specifically, attackers embed triggers into training samples, creating so-called poisoned samples, and strategically align their representations with those of target class samples in the latent space (e.g., through trigger optimization), causing the CL model to be backdoored during the pretraining phase. This backdoor can then be activated either in the encoder or in downstream task models, posing severe security concerns. Nonetheless, data-poisoning-only backdoor attacks in CL are more challenging compared to those in supervised learning, mainly because attackers lose the ability to manipulate annotations, making backdoor injection more difficult in CL \cite{zhang2024data}. In addition, attackers cannot intervene in the pretraining process of CL models or modify the loss function. Consequently, although several data-poisoning-only backdoor attacks have been proposed to date and their feasibility demonstrated, their generalization and robustness remain underexplored.

\begin{table*}[htbp]
\centering
\caption{Limitations of Data-Poisoning-Only CL Backdoor Attacks.  In the ``CL Models" column, black indicates models evaluated in the original works, \textcolor{blue!80}{blue} denotes additional models we evaluated, and \textcolor{red!80}{red} marks omitted original settings (replacing MoCo \cite{he2020momentum} with MoCo v2 \cite{chen2020improved}).}
\resizebox{0.9\linewidth}{!}{
\begin{tabular}{l|l|ccc}
\hline
\multirow{2}{*}{\textbf{Attacks}}    & \multirow{2}{*}{\textbf{CL Models}}  & \multicolumn{3}{c}{\textbf{Sensitivity to}}                                                                  \\ \cline{3-5} 
                                     &                                      & \multicolumn{1}{c|}{\textbf{Downstream Tasks}} & \multicolumn{1}{c|}{\textbf{Datasets}} & \textbf{CL Models} \\ \hline
SSL-Backdoor (CVPR'22) \cite{saha2022backdoor}                & \textcolor{blue!80}{SimCLR}, MoCo v2, BYOL, \textcolor{blue!80}{SimSiam}       & \multicolumn{1}{c|}{Low}                       & \multicolumn{1}{c|}{High}              & Low                \\
PoisonedEncoder (USENIX Security'22) \cite{liu2022poisonedencoder} & SimCLR, \textcolor{blue!80}{MoCo v2}, \textcolor{blue!80}{BYOL}, \textcolor{blue!80}{SimSiam}, \textcolor{red!80}{MoCo} & \multicolumn{1}{c|}{High}                      & \multicolumn{1}{c|}{-}                 & -                  \\
CTRL (ICCV'23) \cite{li2023embarrassingly}                        & SimCLR, \textcolor{blue!80}{MoCo v2}, BYOL, SimSiam       & \multicolumn{1}{c|}{-}                         & \multicolumn{1}{c|}{High}              & High               \\
BLTO (ICLR'24) \cite{sun2024backdoor}                        & SimCLR, \textcolor{blue!80}{MoCo v2}, BYOL, SimSiam, \textcolor{red!80}{MoCo} & \multicolumn{1}{c|}{-}                         & \multicolumn{1}{c|}{Low}               & Low                \\
CorruptEncoder (CVPR'24) \cite{zhang2024data}             & SimCLR, MoCo v2, \textcolor{blue!80}{BYOL}, \textcolor{blue!80}{SimSiam}       & \multicolumn{1}{c|}{High}                      & \multicolumn{1}{c|}{-}                 & -                  \\
NA (ICCV'25) \cite{NA}                         & SimCLR, MoCo v2, BYOL, SimSiam       & \multicolumn{1}{c|}{-}                         & \multicolumn{1}{c|}{Low}               & Low                \\ \hline
\end{tabular}
}
\label{tab:limitations}
\end{table*}

\noindent\textbf{Reevaluation on Data-Poisoning based CL Backdoor.} To systematically assess the generalization and robustness of existing data-poisoning-only backdoor attacks in CL to understand potential caveats or limitations, we establish a unified experimental framework covering six to-date representative methods (SSL-Backdoor \cite{saha2022backdoor}, PoisonedEncoder \cite{liu2022poisonedencoder}, CTRL \cite{li2023embarrassingly}, BLTO \cite{sun2024backdoor},  CorruptEncoder \cite{zhang2024data}, and NA \cite{NA}) across diverse datasets and CL models. These methods span distinct trigger paradigms: pattern-based, static perturbation, and dynamic optimization, covering explicit pixel manipulation to covert feature-space interference. Upon our systematic reproduction and analysis,\autoref{tab:limitations} summarizes the key limitations of six representative data-poisoning-only backdoor attacks.

Our comprehensive evaluation exposes several critical limitations. First, regarding reliance on downstream task priors, PoisonedEncoder and CorruptEncoder exhibit high sensitivity, making them fundamentally unsuitable for encoder-level backdoor injection. SSL-Backdoor shows low sensitivity, which still leads to noticeably reduced attack success rates. In contrast, CTRL, BLTO, and NA are independent of downstream tasks, maintaining effective backdoor performance at the encoder level. Second, concerning dataset and CL model sensitivity, SSL-Backdoor is highly sensitive to dataset resolution, while CTRL is sensitive to both datasets and CL models. Note that BLTO and NA exhibit low sensitivity to datasets and CL models, demonstrating strong cross-model and cross-dataset generalization.
Nevertheless, we observe that despite these noteworthy limitations, the injected trigger consistently enforces representation \textit{similarity among poisoned samples regardless of the samples' original semantic class}. That is, trigger-carrying samples exhibit a high density, which is independent of the original class of the clean image upon which the trigger is attached. As a flip side, this observation inspires us to turn such attacks into an asset; we may repurpose data-poisoning-only backdoor as distinguishable carriers for CL dataset watermarking, thus protecting the dataset's Intellectual Property (IP).

\noindent\textbf{Underexplored CL Dataset IP Protection.} Large-scale CL datasets constructed through web crawling often operate in legal gray areas concerning IP protection. Technology companies, including Meta, OpenAI, and Apple, now face multiple class-action lawsuits alleging unauthorized use of copyrighted content from shadow libraries such as Books3, Bibliotik, and LibGen for training large language models \cite{bairdholm2025apple, lexology2025meta}. The need for dataset IP protection in CL is therefore becoming increasingly important. 
However, existing IP protection solutions remain inadequate. On one hand, many watermarking schemes designed for CL \cite{SSLGuard, SSL-WM, Watermarking} are model-level approaches, embedding backdoors by modifying the model’s loss function during training, under the assumption that watermark injection can \textit{control or intervene in the training pipeline}. In dataset IP protection, however, the dataset owner typically has no access to the training process and cannot control it to facilitate watermark embedding (e.g., by altering the loss function), rendering such model-level watermarking schemes impractical.

To our knowledge, research on dataset IP protection in CL remains strikingly sparse. The few existing studies are all based on dataset fingerprinting---a reactive mechanism (detailed in \autoref{sec:relatedCL_IP}), which suffers from inherent limitations: its verification signal lacks a security key owned by the dataset owner, leaving it vulnerable to false ownership claims \cite{shao2025fitprint}. Most existing watermarking schemes are tailored for supervised learning, where watermarks are tightly coupled with data labels. By contrast, CL derives supervisory signals from the data itself without relying on manual annotations. This fundamental difference renders label-dependent watermarking schemes ineffective. These gaps lead to a research question: \textit{How can we design a watermarking scheme that enables reliable dataset traceback in the absence of labels under CL, without access to or control over the pretraining or downstream task pipelines?}

\noindent\textbf{CL Backdoor for Dataset IP protection.} Nonetheless, in supervised learning, the data-poisoning-only backdoor has been widely used as a proactive dataset IP protection approach \cite{PointNCBW, EntropyMark}. Adopting this concept, this work then presents the first systematic investigation into repurposing data-poisoning-only CL backdoor attacks as its dataset watermarking methods.

However, a watermarked CL encoder is expected to serve diverse downstream tasks, whose data distributions may differ from the pretraining dataset. As a result, relying solely on encoder output verification becomes ineffective under domain shifts or when only the output of the downstream model/classifier is exposed. To address it, we propose a multi-level watermarking scheme that enables dataset ownership verification at three levels (encoder features and downstream soft or hard labels), ensuring robustness to varying downstream scenarios. The implementation involves a two-stage process. During watermark embedding, the dataset owner generates imperceptible watermark patterns using existing data-poisoning-only backdoor attacks and injects a small fraction of such watermark samples into the released dataset. When an unauthorized party trains a CL model on this protected dataset, the model naturally learns to associate these patterns with specific latent representations. For ownership verification, the dataset owner performs black-box verification by querying the suspect encoder or downstream model with inputs containing the watermark pattern. A statistically significant discrepancy between the model’s responses to clean and watermark queries serves as evidence of dataset usage, all without requiring access to the model’s training process or internal parameters.

\noindent\textbf{Contributions.} Our main contributions are threefold:
\begin{enumerate}
    \item We firstly systematically evaluate the attack performance and robustness of six representative data-poisoning-only CL backdoor attacks to date (in \autoref{tab:limitations}) under a unified framework. Our findings reveal their limitations and sensitivities, better understanding of the practicality of these state-of-the-art.

    \item Upon our qualitative and quantitative analysis, four backdoor attacks can be turned into CL dataset watermarking methods by constructively leveraging distinguishable statistical watermark signals, despite low backdoor effect. We design a multi-level black-box dataset watermarking scheme that operates not only on the CL encoder features but also generalizes across unknown downstream tasks by leveraging a unified density-enabled metric.

    \item We conduct comprehensive experiments across four CL model architectures using a multi-dimensional evaluation framework (fidelity, verifiability, robustness). Our results collectively confirm the feasibility of repurposing these four backdoor attacks as effective dataset watermarking, while revealing inherent trade-offs among different methods across evaluation dimensions. It offers practical guidance when choosing them for the CL dataset IP protection.
\end{enumerate}

\section{Preliminary and Related Work}
We first provide the preliminary of CL, followed by related work on CL backdoor attacks and datasets IP protection.

\subsection{Contrastive Learning}
Self-Supervised Learning (SSL) primarily trains models through three interconnected approaches. The first is predictive reconstruction, which includes masked language modeling (e.g., BERT \cite{Bert}) and image patch inpainting (e.g., Context Encoders \cite{ContextEncoders}) to recover deliberately obscured textual segments or visual regions, thereby capturing local and global structures. The second is Contrastive Learning (CL) \cite{chen2020simple, he2020momentum, chen2020improved, chen2021exploring, grill2020Bootstrap}, which optimizes the geometry of the representation space by pulling augmented positive pairs together and pushing negative pairs apart, amplifying inter-sample discriminability. The third is generative modeling (e.g., generative adversarial networks \cite{GAN} or variational autoencoders \cite{VAE}), which implicitly or explicitly reconstructs input data to fit its distribution. Notably, with its ability to learn more discriminative and disentangled visual representations and to generalize exceptionally well across diverse downstream tasks, CL exhibits significant advantages in computer vision.

Traditional explicit CL optimizes representation learning through positive pair agreement maximization and negative pair dissimilarity minimization. SimCLR~\cite{chen2020simple} achieves effective feature learning via intensive within-batch comparisons powered by large batch sizes and sophisticated augmentation pipelines, albeit at substantial computational cost. Note that MoCo~v2~\cite{chen2020improved} is an enhanced version of MoCo~\cite{he2020momentum}, improving CL by maintaining a dynamic queue of negative samples and generating stable negative sample representations through a momentum update encoder. Their general loss function can be defined as

{\small
\begin{equation}
\mathcal{L}=-\frac{1}{N}\sum_{i=1}^{N}\log \frac{\exp(\frac{\text{sim}(z_i^1,z_i^2)}{\tau})}{\exp(\frac{\text{sim}(z_i^1,z_i^2)}{\tau})+\sum_{k\in\mathcal{N}_i}\exp(\frac{\text{sim}(z_i^1,z_k)}{\tau})},
\end{equation}
}
where $z_i^1$ and $z_i^2$ are the projection features of two enhanced views of the same data sample. Here, $\mathcal{N}_i$ is a negative sample set, which means different samples in the same batch in SimCLR, that is, $\mathcal{N}_i={z_j^1, z_j^2|j\neq i}$. In MoCo v2, $\mathcal{N}_i$ represents negative samples in the history queue. The temperature coefficient $\tau$ regulates the sharpness of the similarity distribution.

Beyond these explicit contrastive methods in CL that rely on negative samples, several approaches have emerged that eliminate the need for negative pairs. BYOL~\cite{grill2020Bootstrap} employs an asymmetric architecture with online and target networks, where the online network learns to predict the target network's output representations of different augmented views of the same image. The target network is updated via an exponential moving average of the online network parameters, creating a self-bootstrapping mechanism that prevents representation collapse without negative pairs. Similarly, SimSiam~\cite{chen2021exploring} eliminates negative sampling by focusing solely on maximizing agreement between augmented views of identical instances through innovative stop-gradient operations and a symmetric negative cosine similarity loss
\begin{equation}
\mathcal{L}=-\frac{1}{2}(\frac{p_1}{||p_1||}\cdot\frac{z_2}{||z_2||}+\frac{p_2}{||p_2||}\cdot\frac{z_1}{||z_1||}). 
\end{equation}
Both approaches demonstrate that high quality representations can be learned without explicit negative contrast, with SimSiam being particularly suited for lightweight deployment scenarios due to its simpler architecture.

\subsection{CL Backdoor Attack}
Unlike backdoor attacks in supervised learning, CL typically utilizes unlabeled training data. This fundamental difference means adversaries cannot implement backdoor injections by manipulating data labels. However, poisoning remains feasible through the strategic trigger creation in training samples. Existing backdoor attacks in CL can be classified into two categories based on their dependency on controlling the encoder's training process.

\textbf{Data-Poisoning-Only}. In CL, an encoder or the so-called pretrained model typically acquires generic feature representations during pretraining. Attackers exploit this phase by injecting poisoned data to embed backdoor behaviors through feature encoding permanently. The implanted backdoor can be migrated to any downstream task, does not depend on a specific CL model. Carlini \textit{et al.} \cite{carlinipoisoning} pioneer backdoor injection via toxic image-text pairs in multimodal CL, while Saha \textit{et al.} \cite{saha2022backdoor} propose SSL-Backdoor and achieve similar objectives using pixel patches. The geometrically consistent nature of pixel patches enables implicit learning through multiple augmented views, as their fixed patterns remain detectable despite cropping or enhancement. PoisonedEncoder \cite{liu2022poisonedencoder} directly splices target-class images with inputs, leveraging contrastive loss to enforce feature similarity and establish backdoor associations. CorruptEncoder \cite{zhang2024data} enhances this by using localized target-class objects rather than full images during trigger construction, improving attack success rates. Note that CTRL \cite{li2023embarrassingly} implements stealthier attacks through augmentation-invariant spectral perturbations that overlap with natural feature variations in target classes, causing representation invariance to entangle benign and backdoor features. Meanwhile, BLTO \cite{sun2024backdoor} employs bi-level optimization: the inner loop simulates CL on poisoned data, while the outer loop updates a backdoor generator to minimize feature distance between triggers and target embeddings. More recently, Chen \textit{et al.} \cite{NA} propose Noisy Alignment (NA), which suppresses noise components via optimized image layout.

\textbf{Encoder-Controlled-Poisoning.} This attack paradigm diverges from data poisoning only approaches by \textit{requiring either architectural knowledge or control} of the CL encoder's training mechanics (e.g., loss functions, optimization strategies, or model parameters). Through dynamic parameter optimization and customized training protocols, attackers achieve enhanced backdoor persistence. BadEncoder \cite{jia2022badencoder} pioneers gradient-driven backdoor embedding, simultaneously optimizing trigger effectiveness and model utility via dual loss minimization. GhostEncoder \cite{GhostEncoder} employs steganographic techniques to implant hidden triggers in benign images, subsequently fine-tuning pretrained encoders to internalize backdoor mappings. Building on Bayesian principles, Liang \textit{et al.}'s BadCLIP \cite{LiangBadCLIP} establishes cross-modal backdoor alignment by minimizing distances between visual triggers and textual targets in embedding space. Bai \textit{et al.}'s parallel BadCLIP \cite{BaiBadCLIP} implementation extends this to multimodal CL, deploying trigger-aware context generators and adaptively optimized triggers to jointly corrupt image-text feature spaces. This bimodal synchronization amplifies target-class feature similarity during trigger activation, substantially improving attack success rates. Addressing detection vulnerabilities, Tao \textit{et al.} \cite{tao2024distribution} confront the feature-space discreteness of conventional backdoor samples through Kernel Density Estimation (KDE) \cite{KDE} with sliced-Wasserstein distance \cite{SlicedWassersteinDistances}. Their method enforces distributional congruence between poisoned and benign samples in feature space, effectively concealing backdoor patterns from clustering-based defenses.

We consider evaluating all six data-poisoning-only backdoor attacks (as in \autoref{tab:limitations}) in CL that operate exclusively on image-only modality without textual label manipulation. While Carlini \textit{et al.} \cite{carlinipoisoning} pioneered backdoor injection via poisoned image-text pairs, their approach inherently modifies the textual supervision signal, thus being analogous to label manipulation in supervised learning and comparatively less challenging than purely unsupervised image-based poisoning. We therefore exclude this multimodal setting.

\subsection{CL Dataset IP Protection}\label{sec:relatedCL_IP}
The growing dependence of artificial intelligence on high quality training data has made datasets IP protection a crucial challenge for ensuring model compliance and responsible development. In supervised learning, existing IP protection schemes commonly rely on model predictions and human-provided annotations. Typical methods include backdoor based watermarking \cite{tang2023did, li2022untargeted, li2020open}, decision boundary auditing \cite{maini2021dataset, christopher2021label}, and model behavior analysis \cite{sablayrolles2020radioactive, christopher2021label}, all of which utilize the model’s class probabilities or output labels to verify datasets usage.

In contrast, CL learns general-purpose feature representations directly from unlabeled data without depending on manual annotation signals. As a result, the conventional protection methods mentioned above are not directly applicable to the CL setting. Recent research on datasets IP protection in CL has largely focused on datasets fingerprinting. The underlying premise is that a dataset used to train a model leaves identifiable statistical traces (or ``fingerprints") in the model’s internal representations or behavioral patterns. By extracting and analyzing these fingerprints, one can trace whether a particular dataset is used in training. For example, Dziedzic \textit{et al.} \cite{dziedzic2022dataset} propose a dataset inference method that compares the representation distributions of encoder outputs on training versus test data. A significantly higher log likelihood for a given dataset suggests the model is trained on it. Xie \textit{et al.} \cite{xie2025dataset} introduce a datasets IP verification method for CL pretrained models, analyzing differences in encoder representations between trained and new data to define a ``contrastive relational gap", which helps determine datasets usage. Their work shows that representations of trained data exhibit stronger self-relations (greater similarity between different augmentations of the same sample) and more stable binary relations (less change in inter-sample similarity after augmentation).

Although fingerprinting can trace datasets usage in CL, its verification signal originates from general model data interaction patterns and lacks a security key that is exclusively owned and actively controlled by the dataset owner. This opens the possibility for adversarial parties to employ the same method to falsely claim ownership, leading to disputes over IP attribution \cite{shao2025fitprint}. To address this limitation, this paper proposes repurposing data-poisoning-only based CL backdoor attacks as datasets watermarking methods. The key distinction from fingerprinting lies in the ownership and control of the verification key. In the watermarking scheme, the dataset owner holds a unique and non-replicable method for generating trigger patterns. Prior to datasets release, the owner injects poisoned samples containing these triggers into the dataset, thereby embedding watermarks. During verification, the owner queries a suspect model’s Application Programming Interface (API) with trigger embedded samples and checks for an expected anomalous response. This process enables label free IP verification and establishes an active, controllable, and legally substantive mechanism for datasets IP protection.

\section{Evaluating Data-Poisoning-Only based CL Backdoor Attacks}

\subsection{Data-Poisoning-Only based CL Backdoor Attacks}
\label{opt}
In the following, we first briefly introduce the implementation principles of data poisoning-only backdoor attacks in CL. We then systematically evaluate their performance and robustness through an end-to-end experiment setting, and consequently analyze their feasibility for dataset IP watermarking from both quantitative and qualitative perspectives. 

\textbf{SSL-Backdoor} \cite{saha2022backdoor} aims to inject a hidden backdoor into a CL model such that a downstream classifier built on top of it misclassifies any test image containing a specific trigger patch as a predefined target class, while maintaining normal performance on clean inputs. To achieve this, the attacker selects a target class that exists in the unlabeled pretraining dataset $\mathcal{D}^{\rm PT}$ and also typically appears in the downstream dataset $\mathcal{D}^{\rm DT}$. The attacker then poisons \(\mathcal{D}^{\rm PT}\) by pasting a small trigger patch (e.g., a colored square) onto a fraction of the images belonging to that target class and injects these poisoned images into the pretraining dataset. Subsequently, an exemplar‑based CL method is trained on the poisoned $\mathcal{D}^{\rm PT}$. These methods rely on an inductive bias that two different random augmentations of the same image should produce similar embeddings. Due to the rigid appearance and low variation of the trigger, the model learns a strong implicit detector for it. Since the trigger co‑occurs only with images of the target class during pretraining, the model associates the trigger’s visual features with that class. After pretraining, a linear classifier is trained on top of the frozen CL encoders using a small labeled subset of $\mathcal{D}^{\rm DT}$. At test time, the classifier performs normally on clean images. However, when the same trigger is pasted onto any test image (even from a different class), the model classifies it as the target class, thereby achieving a successful backdoor attack.

\textbf{PoisonedEncoder}~\cite{liu2022poisonedencoder} shares a similar attack objective on the downstream task as SSL-Backdoor, but with a different technique. The attack can be understood and explained as a formulated bi-level optimization during the victim encoder training (access to the victim encoder is not needed, and the solution is detailed shortly). The outer objective maximizes the feature similarity between each target input and reference inputs from the target class, while the inner objective corresponds to the standard CL process that trains an encoder. Formally, let $x_{ti}$ denote a target input for the $i$-th target of the $t$-th downstream task dataset $\mathcal{D}^{\rm DT}$, $x_r \in X_{y_{ti}}$ a reference input from the set of data from the target class $y_{ti}$, and $\theta$ the encoder parameters. The attack objective is:
\begin{equation}
    \max_{X_p} \frac{1}{S} \sum_{t=1}^{T}\sum_{i=1}^{k_t}\sum_{x_r \in X_{y_{ti}}} \mathcal{L}_{\text{sim}}(x_{ti}, x_r; \theta^*(X_c \cup X_p)),
\end{equation}
\begin{equation}
    \text{s.t. } \theta^*(X_c \cup X_p) = \arg\min_{\theta} \mathcal{L}_{\text{CL}}(X_c \cup X_p; \theta),
\end{equation}
where $T$ is the number of target downstream tasks, $k_t$ is the number of target inputs for task $t$, $X{y_{ti}}$ is the set of reference inputs from the target class $y_{ti}$, $|X_{y_{ti}}|$ denotes the size of that set, $S = \sum_{t=1}^{T}\sum_{i=1}^{k_t}|X_{y_{ti}}|$ is the total number of (target input, reference input) pairs used for normalization, $\mathcal{L}_{\text{sim}}$ is the cosine similarity between the feature vectors of $x_{ti}$ and $x_r$ (the outer objective aims to maximize this similarity), $\mathcal{L}_{\text{CL}}$ is the contrastive loss used to pretrain the encoder, $X_c$ is the set of clean unlabeled pretraining dataset $\mathcal{D}^{\rm PT}$, and $X_p$ is the set of poisoning inputs crafted by the attacker.

Solving this bi-level problem directly is challenging due to the inaccessibility of gradients, the black-box nature of the encoder, and the lack of knowledge about the clean dataset. To address these challenges, PoisonedEncoder leverages the random cropping augmentation intrinsic to CL. Specifically, the attacker constructs poisoning inputs by spatially concatenating a target input with a reference input from the target class using four geometric configurations: top-bottom, bottom-top, left-right, and right-left. During pretraining, the random cropping operation may separately capture the target and reference regions from the same composite image. Since CL enforces that two randomly cropped views from the same image have similar feature representations, the encoder is implicitly driven to align the feature of the target input with that of the reference input. This mechanism effectively fulfills the outer optimization objective without requiring knowledge of the clean dataset, encoder architecture, or loss function.

\textbf{CTRL}~\cite{li2023embarrassingly} exploits the representation invariance property of CL by constructing an augmentation‑insensitive trigger in the frequency domain and embedding it into target class samples, thereby entangling the representations of poisoning and target class samples that the adversary assumes to possess from the unlabeled pretraining dataset $\mathcal{D}^{\rm PT}$ in the feature space. Specifically, the attacker applies a low-frequency perturbation to the chrominance channels in the YCbCr color space via the Discrete Cosine Transform (DCT) \cite{DCT}. This trigger is visually imperceptible and robust to common augmentations such as random cropping and color jittering. Let $x$ be a clean target class sample and $r$ be the trigger. The poisoning sample $x_* = x \oplus r$ can be approximated as a linear mixture in the feature space:
\begin{equation}
    f(x_*) = (1-\alpha)f(x) + \alpha f(r),
\end{equation}
where $\alpha \in (0,1)$ is the mixing weight. During CL, the model forces different augmented views of the same sample to be close in the feature space. For a poisoning sample $x_*$ and its augmented view $x_*^+$, the cosine similarity expands to:
\begin{multline}
    f(x_*)^\top f(x_*^+) = (1-\alpha)^2 f(x)^\top f(x^+) + \alpha^2 f(r)^\top f(r^+) \\
    + \alpha(1-\alpha)\left(f(x)^\top f(r^+) + f(r)^\top f(x^+)\right),
\end{multline}
where the cross term $\alpha(1-\alpha)(\cdot)$ aligns the trigger $r$ with the target class sample $x$, gradually entangling their representations. As a result, any input embedded with the trigger is misclassified into the predefined target class by the downstream classifier.

\textbf{BLTO}~\cite{sun2024backdoor} utilizes a bi-level optimization (i.e., inner and outer optimizations) designed to generate triggers that can adapt to the unique mechanisms of CL. The inner optimization simulates the CL training process of a victim surrogate model on a poisoning dataset. More specifically, given a clean pretraining dataset $\mathcal{D}^{\rm PT}$, the adversary is assumed to hold reference data $\mathbf{x}_r$ sampled from the target class, which is assumed to be contained within $\mathcal{D}^{\rm PT}$; despite the overall pretraining dataset being unlabeled, the adversary possesses labels for this small subset of target class data. Using the current backdoor generator $g_{\psi}(\cdot)$, the adversary constructs the poisoned pretraining dataset $\mathcal{D}^{\rm PT}_b$ by combining the clean data with backdoor-transformed reference examples, i.e., $\mathcal{D}^{\rm PT}_b = \mathcal{D}^{\rm PT} \cup g_{\psi}(\mathbf{x}_r)$. A surrogate feature extractor $f_{\theta}$ with the same architecture as the target CL model is then trained on $\mathcal{D}^{\rm PT}_b$ by minimizing the standard contrastive loss $\mathcal{L}_{\text{CL}}$:
\begin{equation}
    \theta = \arg\min_{\theta} \mathbb{E}_{\mathbf{x} \in \mathcal{D}^{PT}_b} \mathcal{L}_{\text{CL}}(\mathbf{x}; \theta).
\end{equation}
This step mimics the backdoored learning dynamics under poisoned data, by incorporating the effects of data augmentation and uniformity regularization. The outer optimization updates the backdoor generator to maximize attack effectiveness. Its objective is to maximize the similarity between the poisoning data and the target class reference data in the embedding space:
\begin{equation}
    \max_{\psi} \mathbb{E}_{\mathbf{x} \in \mathcal{D}^{PT}, t_1, t_2 \in \mathcal{T}} \left[ S\left( f_{\theta}(t_1(g_{\psi}(\mathbf{x}))), f_{\theta}(t_2(\mathbf{x}_r)) \right) \right],
\end{equation}
where $\mathcal{T}$ denotes the set of augmentations and $S(\cdot, \cdot)$ is a similarity measure. By fixing $\theta$ during inner updates and optimizing $\psi$ in outer updates, and alternating these steps iteratively, BLTO produces a trigger that not only deceives the surrogate model but also generalizes effectively to unseen victim CL pipelines, attributing to transferability, thereby achieving high attack success rates and robustness.

\textbf{CorruptEncoder}~\cite{zhang2024data} exploits the random cropping mechanism to align the trigger with target class objects from the downstream task dataset $\mathcal{D}^{\rm DT}$ in the feature space. The attacker embeds a small set of reference objects and a trigger into background images following a theoretically optimal layout: the reference object is placed at a corner, the trigger near the center of the remaining area, and the background image is sized to approximately twice the reference object. This layout maximizes the probability that two randomly cropped views capture only the reference object and only the trigger, respectively, formalized as:
\begin{equation}
    p = \frac{1}{S} \int_{s \in (0, S]} p_1(s) p_2(s) \, ds,
\end{equation}
where $p_1(s)$ and $p_2(s)$ denote the probabilities of a crop covering the reference object or the trigger. During pretraining, the encoder is optimized to maximize feature similarity between two cropped views, thereby learning to map the trigger and reference object into similar representations. Consequently, a downstream classifier built on this encoder classifies trigger-carrying images as the target class while preserving accuracy on clean inputs. To enhance attack robustness, CorruptEncoder+ introduces support reference images and jointly optimizes the similarity between the trigger and reference object, and between the reference object and its class centroid:
\begin{equation}
    \max_{\mathcal{D}_p} \left[ S_C(f_o, f_e) + \lambda \cdot S_C(f_o, f_{cls}) \right],
\end{equation}
which improves the discriminability of reference objects in the feature space and increases attack success.

\textbf{Noisy Alignment (NA)}~\cite{NA} aims to inject a small set of carefully crafted composite poisoned images into the pretraining dataset $\mathcal{D}^{\rm PT}$ so that the resulting encoder will misclassify any test sample carrying a trigger $\mathbf{p}$ as a predefined target class $t$ in the downstream dataset $\mathcal{D}^{\rm DT}$ (where class $t$ exists both among the unlabeled images in $\mathcal{D}^{\rm PT}$ and as a class in $\mathcal{D}^{\rm DT}$). 
To achieve this, the attacker first collects a shadow image set $\mathcal{D}_{\rm shadow}$ (arbitrary classes) and a reference image set $\mathcal{D}_{\rm ref}$ (from the target class $t$). For each pair of a shadow image $\mathbf{x}_s \in \mathcal{D}_{\rm shadow}$ and a reference image $\mathbf{x}_r \in \mathcal{D}_{\rm ref}$, the attacker embeds the trigger $\mathbf{p}$ into $\mathbf{x}_s$ to obtain $\mathbf{x}_s \oplus \mathbf{p}$, and then composes a composite image $\overline{\mathbf{x}_s,\mathbf{x}_r}$ according to the theoretically optimal layout: the reference image is placed at the top-left corner $(0,0)$ of the canvas, the poisoned shadow image is placed at the horizontal center $(c_w/2, 0)$, the trigger is centered within the shadow region, and the canvas size is set to width $c_w = 2r_l$ and height $c_h = r_l$ (where $r_l$ is the side length of each image), with the reference and shadow regions tightly adjacent and non-overlapping. 
This layout maximizes the joint probability that, under the random cropping augmentations $T_1, T_2 \stackrel{i.i.d.}{\in} \mathcal{T}$ used in CL, the two cropped views $\nu_1 = T_1(\overline{\mathbf{x}_s,\mathbf{x}_r})$ and $\nu_2 = T_2(\overline{\mathbf{x}_s,\mathbf{x}_r})$ satisfy $\mathbf{p} \subseteq \nu_1$, $\nu_1 \subseteq (\mathbf{x}_s \oplus \mathbf{p})$, $\nu_2 \subseteq \mathbf{x}_r$, and $\nu_1 \cap \nu_2 = \emptyset$, i.e., $\Pr(\mathbf{p}\subseteq \nu_1,\ \nu_1\subseteq(\mathbf{x}_s\oplus\mathbf{p}),\ \nu_2\subseteq\mathbf{x}_r,\ \nu_1\cap\nu_2=\emptyset)$. 

During pretraining on $\mathcal{D}^{\rm PT}$, the contrastive model treats $\nu_1$ and $\nu_2$ as a positive pair, forcing the encoder to align the trigger features with the reference features of the target class while compressing irrelevant noisy features orthogonal to the reference direction. Consequently, after such poisoned pretraining, any test input containing the same trigger $\mathbf{p}$ will be misclassified as the target class $t$ by a downstream linear classifier trained on $\mathcal{D}^{\rm DT}$.



\begin{figure*}[htbp]
    \centering
    \includegraphics[width=0.98\linewidth]{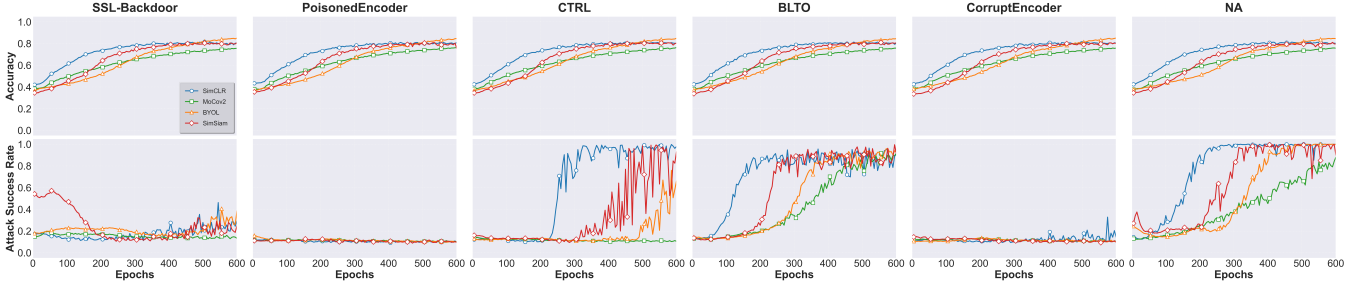} 
    \caption{Comparison of model accuracy and attack success rate under different backdoor attacks and model structures on CIFAR10.}
    \label{fig:cifar10}
\end{figure*}

\subsection{Backdoor Performance Evaluation}
\label{backdoor}
Under a unified experimental framework, we now evaluate the backdoor effectiveness of each method from the perspective of CL encoders. We adopt CIFAR10 \cite{cifar2009} as the pretraining dataset, set the target class to ``truck”, and use a small poisoning rate of 1\%. All methods are evaluated on all four CL models with ResNet18~\cite{ResNet} as the backbone, namely SimCLR~\cite{chen2020simple}, Moco v2~\cite{chen2020improved} , BYOL~\cite{grill2020Bootstrap} and Simsiam~\cite{chen2021exploring}, for 600 epochs.

Results are presented in \autoref{fig:cifar10}. In terms of model accuracy (ACC) for clean image inputs, none of the backdoor attacks significantly compromises the utility of the pretrained encoders; after 600 epochs, all models achieve accuracy between 75\% and 85\%, with most around 80\%. 

Regarding attack success rate (ASR), SSL-Backdoor achieves only moderate ASR, ranging from approximately 13\% on MoCo v2 to 39\% on BYOL, indicating limited effectiveness. PoisonedEncoder fails to inject a successful backdoor at the encoder level, as its ASR never exceeds the random baseline (around 10\%) across all model architectures. CTRL is highly sensitive to the model architecture: while its ASR exceeds 90\% on SimCLR (97.9\%) and SimSiam (93.2\%), it reaches about 66\% on BYOL and completely fails on MoCo v2, where ASR remains at the random guessing level (approximately 10\%). In contrast, BLTO consistently achieves the best performance across all four model architectures, with ASR steadily exceeding 90\% (ranging from 90.8\% on SimCLR to 92.8\% on BYOL) and converging after 500 epochs. 
Similar to PoisonedEncoder, CorruptEncoder also fails to achieve a successful backdoor injection, with ASR staying around 10\% on all models. Finally, NA attains high ASR, exceeding 88\% on all models and reaching nearly 100\% on BYOL and SimCLR. 

The underlying reason for the failure of PoisonedEncoder and CorruptEncoder is that both methods rely on the assumption that the target class does not appear in the pretraining dataset, which allows the model to uniquely associate the trigger with that class. When the target class is changed to an existing class in the pretraining dataset (i.e., ``truck”), the presence of a large number of clean samples from that class interferes with the stable learning of the backdoor association, ultimately preventing successful injection. Therefore, in the following experiments on additional datasets, we exclude PoisonedEncoder and CorruptEncoder due to their inherent limitations.



\begin{figure*}[htbp]
    \centering
    \includegraphics[width=0.8\linewidth]{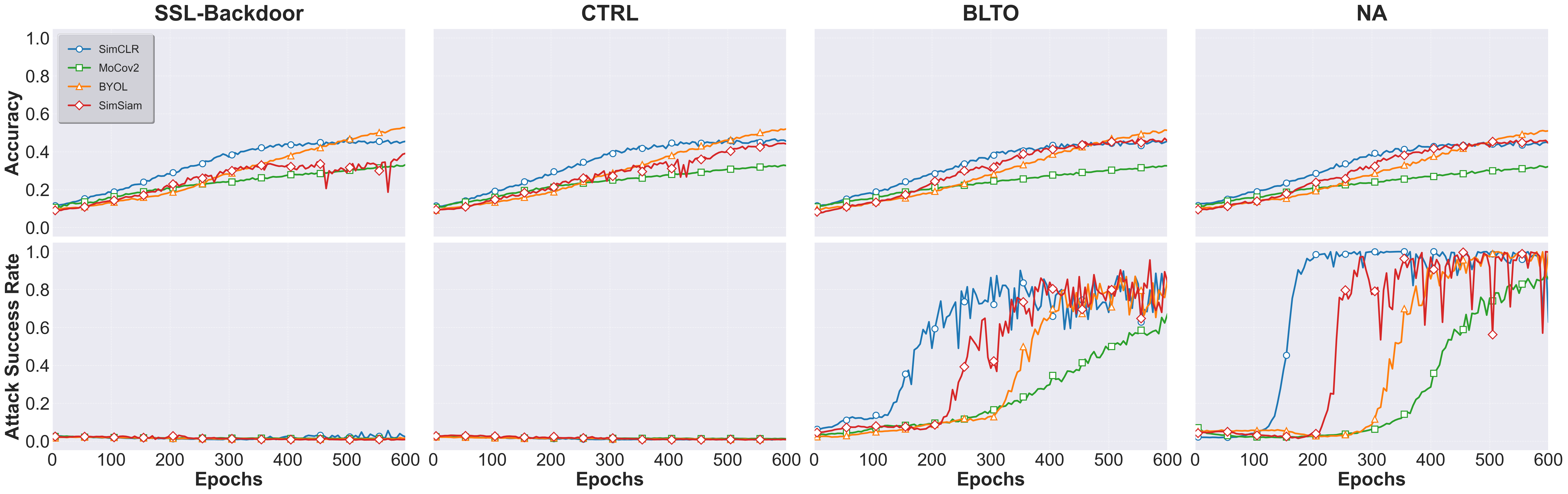} 
    \caption{Comparison of model accuracy and attack success rate under different backdoor attacks and model structures on ImageNet100.}
    \label{fig:imagenet100}
\end{figure*}

We further evaluate the performance of SSL-Backdoor, CTRL, BLTO, and NA on the more complex ImageNet100 \cite{imagenet} dataset. Experimental results in \autoref{fig:imagenet100} show that SSL-backdoor and CTRL almost completely fail on ImageNet100, with their ASR remaining near the random baseline. They are highly sensitive to data distribution and exhibit limited generalization. BLTO maintains a relatively high ASR on ImageNet100, though the ASR decreases by approximately 10\% compared to that on CIFAR10 and exhibits considerable fluctuation during training. In contrast, NA delivers the most robust performance across all CL frameworks. It maintains a close to 100\% ASR on SimCLR, BYOL, and SimSiam despite noticeable fluctuations, and still achieves around 85\% ASR on MoCo v2.

\subsection{Qualitative Analysis}
Based on the above experimental evaluation, we first note that the six methods exhibit limitations when assessed strictly at the pretrained encoder level, but not the downstream classifier. PoisonedEncoder and CorruptEncoder fail to inject an effective backdoor in the unified setting because their original designs rely on knowledge of downstream-task assumptions. 

Among the remaining methods, SSL-Backdoor and CTRL demonstrate some capability for backdoor injection but suffer from considerable instability across different model architectures and datasets: CTRL shows high variance (e.g., achieving over 90\% ASR on SimCLR but near random on MoCo v2) and completely fails on ImageNet100; SSL-Backdoor achieves only moderate ASR on CIFAR10 (up to 39\% on BYOL) and also nearly fails on ImageNet100. 

In contrast, BLTO and NA show the best and most stable performance. Note that BLTO consistently exceeds 90\% ASR across all architectures on CIFAR10 and maintains a relatively high ASR on ImageNet100 (though with a 10\% drop), while NA achieves near-perfect ASR (over 99\%) on CIFAR10 and remains effective on ImageNet100. 

Despite these limitations, we observe that backdoor behaviors of these methods can be reinterpreted as potential signals for dataset ownership verification. This leads us to the key insight: although originally designed as backdoor attacks, these methods may be repurposed for dataset watermarking in CL.

\noindent\textbf{Re-purpose Feasibility.} We therefore proceed to theoretically analyze the feasibility of adapting these methods for dataset watermarking. From the implementation perspective, the backdoor effects of PoisonedEncoder and CorruptEncoder are primarily designed for assumed known downstream classification tasks, with their triggering mechanisms relying on a target class prespecified by the attacker. For CL dataset IP protection, dataset owners typically cannot anticipate the specific downstream tasks or datasets, nor can they assume that any downstream task will include a particular predefined class. Consequently, these two methods are fundamentally unsuitable for dataset watermarking. 

In contrast, the backdoors injected by SSL-Backdoor, CTRL, BLTO, and NA, are all independent of downstream task classes. They steer poisoned samples toward an existing class within the pretraining dataset itself, meaning that backdoor activation is determined solely by the internal structure of the pretraining dataset, without reliance on any downstream task. Although some of these methods exhibit instability or only moderate ASR across different architectures and datasets, we argue that dataset watermarking does not require a high ASR to be effective. Even a weak or inconsistent backdoor can offer a detectable statistical shift in the model's behavior on triggered inputs, which can suffice as a watermark signal for ownership verification. 

Therefore, we proceed to evaluate all four methods as potential dataset watermarking methods. Once a backdoor is successfully embedded, a model pretrained via CL will map the feature representations of inputs containing the trigger toward the feature region corresponding to the target class. Such anomalous behavior can be regarded as a watermark signal, enabling ownership verification based on an internal class of the pretraining dataset without requiring prior knowledge of downstream tasks. The next section details the complete implementation of repurposing these four backdoor attacks as reliable dataset watermarking verification methods.

\section{Repurposing Backdoor for Watermarking}
This section first defines the threat model of dataset IP protection, followed by the implementation details of repurposing CL backdoor attacks into dataset watermarking methods.

\subsection{Threat Model}
Our threat model considers two entities: Pretraining Dataset Owner and Model Developer.

\begin{itemize}
    \item \textbf{Pretraining Dataset Owner}, is the IP owner. Before public release, the owner embeds watermarks by injecting a small set of watermark data containing specific trigger patterns into the dataset. If the protected dataset is subsequently used to train a CL model, the resulting model will inadvertently inherit the hidden watermark. The owner can later verify ownership by querying the model's API, whether accessing the pretrained encoder or a downstream task model, to detect the presence of the embedded watermark.
    \item \textbf{Model Developer}, is the entity that procures datasets, potentially from web scraping \cite{cdi}, and trains CL models. The developer follows standard training procedures and is typically unaware of the presence of watermarks in the acquired data. After training, the developer may release the pretrained encoder publicly or fine-tune it for various downstream tasks, offering API based access and charging users accordingly.
\end{itemize}

\begin{figure*}[htbp]
    \centering
    \includegraphics[width=0.9\linewidth]{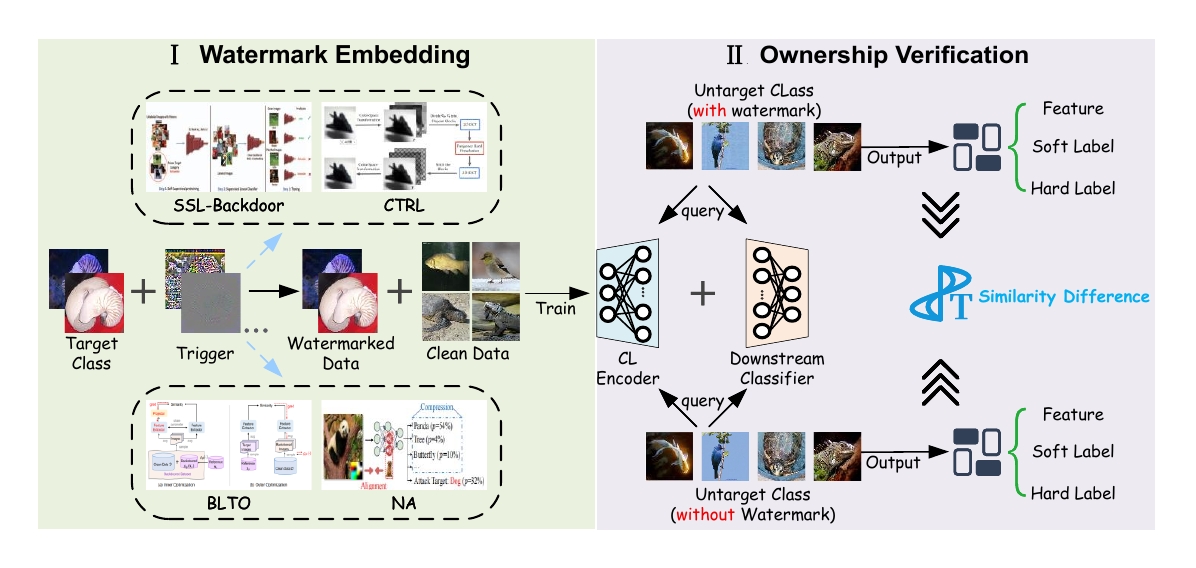} 
    \caption{The process of repurposing feasible data-poisoning-only based backdoor attacks in CL into a datasets watermarking method. The framework consists of two stages: watermark embedding and ownership verification.}
    \label{fig:scheme}
\end{figure*}

\subsection{Repurpose and Implementation for Dataset Watermarking}
Note that CL dataset watermarking has two phases: watermark embedding and ownership verification, with the overall process illustrated in \autoref{fig:scheme}.

\noindent\textbf{Watermark Embedding.} During this phase, the pretraining dataset owner, acting as the defender, embeds an imperceptible watermark (i.e., a backdoor trigger) into the protected dataset. Specifically, similar to the assumption in backdoor attacks, the owner possesses a small subset of samples from a specific class within the otherwise unlabeled CL dataset, and selects that class as the target for watermarking. Leveraging one of the four data-poisoning-only backdoor attacks (SSL-Backdoor, CTRL, BLTO, or NA), the owner then generates poisoned samples by fusing the corresponding trigger into these known target class samples. These poisoned/watermarked samples constitute the watermark set and are mixed with the original clean dataset at a small ratio to form the final released dataset. The backdoor effect inherently ensures the effectiveness of the embedded watermark.

\noindent\textbf{Ownership Verification.} In this phase, when the dataset owner suspects unauthorized use of their dataset, they can initiate a black-box ownership verification operation. They do not need to access the model's internal parameters, but instead perform verification solely through the query API. The verification is performed as follows: The dataset owner, as the verifier, selects a diverse set of clean query samples from non-target classes and generates their corresponding watermark versions using the same trigger. Both the clean and watermark samples are then fed into the suspect model to obtain its outputs, which can be feature representations, soft labels, or hard labels depending on the model's API.

Based on the model outputs, the verifier computes two metrics: the mean cosine similarity $S$ among the outputs of clean samples, and the mean cosine similarity $S'$ among the outputs of watermark samples. Specifically,
\begin{equation}
    S=\frac{2}{n(n-1)}\sum_{i=1}^{n-1}{\sum_{j=i+1}^{n}\cos(x_i,x_j)}
\end{equation}
where $n$ is the number of query samples, and $x_k,k \in \{1,2,...,n\}$ denotes the model output for the $k_{\rm th}$ clean sample.
Similarly,
\begin{equation}
    S'=\frac{2}{n(n-1)}\sum_{i=1}^{n-1}{\sum_{j=i+1}^{n}\cos(x_i',x_j')}
\end{equation}
where $x_k'$is the corresponding watermarked version of $x$.

\noindent\textbf{Rationale.} Our rationale is rooted in the concept of output density. Clean samples are drawn from diverse classes, so their outputs scatter across the representation space with low density, leading to a small $S$. In contrast, even a weak backdoor can create a local density concentration: inputs embedded with the same trigger are mapped to nearby regions in the output space, be it feature vectors, softmax probability vectors, or even hard label distributions, because they share the same trigger-induced perturbation. This density concentration is detectable through cosine similarity, regardless of whether the outputs are features (where similarity directly reflects geometric proximity), soft labels (where similar probability vectors indicate clustered predictions), or hard labels (where identical label assignments produce perfect similarity). Therefore, $S'$ should be significantly higher than $S$ as long as the backdoor exerts any systematic pull on the model's outputs. If the difference $\Delta = S' - S$ is statistically significantly greater (as determined by an appropriate statistical test) than a predefined threshold $\tau$, then it can be inferred based on the principle of low probability events that the model's training datasets contained the specific watermark pattern. This provides a valid foundation for claiming ownership of the dataset. Notably, this verification framework unifies the three output types (feature representations, soft labels, and hard labels that can be represented by a one-hot vector) under a single approach.


\subsection{Evaluation Aspects}
Three aspects are used to evaluate watermark performance: Fidelity, Verifiability, and Robustness.

\noindent\textbf{Fidelity} measures the watermark's impact on both the perceptual quality of the dataset and the utility of models trained on it. A high-fidelity watermark should preserve the semantic content and statistical distribution of the original data with minimal visual degradation. To quantitatively assess perceptual similarity between clean and watermark images, we adopt three complementary metrics: SSIM \cite{SSIM} captures structural consistency, LPIPS \cite{LPIPS} reflects perceptual differences based on deep features, and DreamSim \cite{DreamSim} further aligns with human similarity judgments by incorporating semantic feature fusion. To evaluate model utility, we measure model accuracy from two perspectives: the encoder level and the downstream task level. A high-fidelity watermark ensures that models trained on the watermarked dataset achieve accuracy comparable to those trained on the clean dataset at both levels, thereby preserving the dataset's practical utility.

\noindent\textbf{Verifiability} refers to the ability to correctly confirm dataset ownership through watermarking mechanisms. This aspect requires that the dataset owner can reliably extract predefined signals from models trained on the watermarked (IP) dataset, while such signals remain absent from models trained on clean (non-IP) datasets, thereby ensuring credible ownership attribution. To quantify verifiability, we employ two complementary indicators: the similarity difference $\Delta$ and the statistical P-value. Specifically, $\Delta = S' - S$, where $S'$ is the mean cosine similarity among outputs of watermarked query samples and $S$ is that of clean query samples. A larger $\Delta$ indicates a stronger watermark signal. Based on $\Delta$, we formulate a hypothesis test: the null hypothesis $H_0$ assumes that the suspect model is trained on a non‑IP dataset, i.e., $\Delta_{\text{suspect}} \leq \tau$, while the alternative $H_1$ assumes it is trained on the IP dataset, i.e., $\Delta_{\text{suspect}} > \tau$, where $\tau$ is a predefined threshold. The P‑value quantifies the probability of observing $\Delta_{\text{suspect}}$ under $H_0$ using a one‑sample t‑test. If the P‑value is below the significance level $\alpha = 0.05$, we reject $H_0$ and conclude that the suspect model was trained on the IP dataset (i.e., infringement) \cite{dziedzic2022dataset, xie2025dataset}. High verifiability is characterized by a clear margin in $\Delta$ between IP‑trained and non‑IP‑trained models, as well as low false positive and high true positive rates in the hypothesis test.

\noindent\textbf{Robustness} evaluates the watermark's ability to persist under typical CL applications. Since a pretrained CL encoder is often used as a foundation model for various downstream tasks, a robust watermark should remain detectable after fine-tuning to different tasks, enabling reliable and continuous ownership tracking. We continue to use the similarity difference $\Delta$ as the metric to quantify robustness.

\begin{figure}[htbp]
    \centering
    \includegraphics[width=0.99\linewidth]{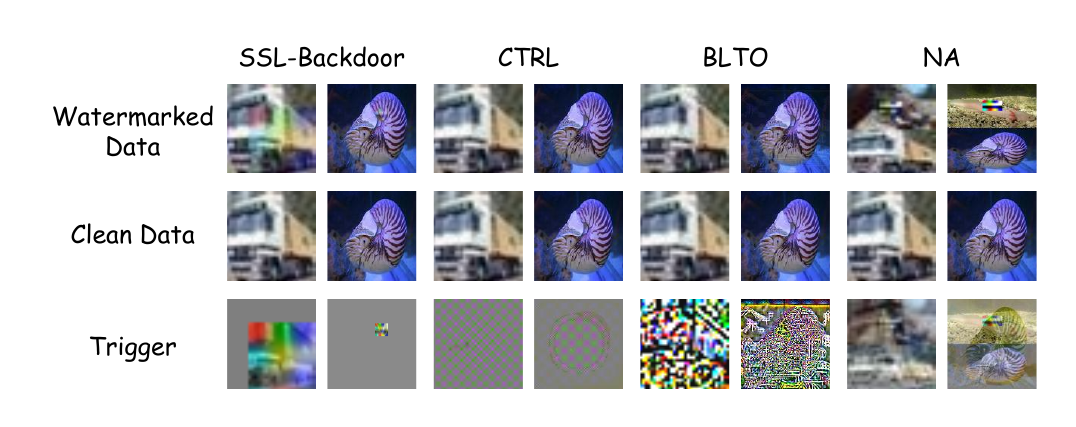} 
    \caption{Watermarked sample example: CIFAR10 (left), ImageNet100 (right).}
    \label{fig:watermark}
\end{figure}

\subsection{Fidelity}
Following the experimental setups of the original backdoor works, we generate and embed watermarks on the CIFAR10 and ImageNet100 datasets, respectively. \autoref{fig:watermark} shows examples of watermark data for the ``truck" class in CIFAR10 and the ``nautilus" class in ImageNet100. It can be observed that both CTRL and BLTO exhibit clear patterns when presented alone, yet become nearly imperceptible once added to the corresponding data. In contrast, SSL-Backdoor and NA, which rely on patch-based triggers, introduce visible pixel blocks in the watermark data. Moreover, NA employs an additional stitching strategy, making its watermark data deviate even more from clean data.

\begin{table}[htbp]
\centering
\caption{Fidelity comparison on CIFAR10 and ImageNet100.}
\resizebox{0.95\linewidth}{!}{
\begin{tabular}{c|cccc}
\hline
\textbf{Datasets}                     & \textbf{Methods}      & \textbf{SSIM $\uparrow$} & \textbf{LPIPS $\downarrow$} & \textbf{DreamSim $\downarrow$} \\ \hline
\multirow{4}{*}{\textbf{CIFAR10}}     & \textbf{SSL-Backdoor} & 0.8815        & 0.2173         & 0.2920            \\
                                      & \textbf{CTRL}         & 0.9537        & 0.0933         & 0.1013            \\
                                      & \textbf{BLTO}         & 0.9526        & 0.0774         & 0.1038            \\
                                      & \textbf{NA}           & 0.1232        & 0.5664         & 0.4255            \\ \hline
\multirow{4}{*}{\textbf{ImageNet100}} & \textbf{SSL-Backdoor} & 0.9913        & 0.0216         & 0.0243            \\
                                      & \textbf{CTRL}         & 0.9928        & 5.83e-5        & 0.0015            \\
                                      & \textbf{BLTO}         & 0.8254        & 0.1243         & 0.1415            \\
                                      & \textbf{NA}           & 0.2235        & 0.6494         & 0.4220            \\ \hline
\end{tabular}
}
\label{tab:fidelity}
\end{table}

\subsubsection{Perceptual Quality} We employ three complementary quantitative metrics, SSIM \cite{SSIM}, LPIPS \cite{LPIPS}, and DreamSim \cite{DreamSim}, to measure the perceptual quality of the watermarked images, with results reported in \autoref{tab:fidelity}.

On CIFAR10, SSL‑Backdoor achieves moderate fidelity. Its patch occupies a relatively large area on low‑resolution $32\times32$ images, causing noticeable but not severe distortion. CTRL and BLTO exhibit significantly better fidelity, both with SSIM above 0.95, making their watermarks nearly imperceptible. NA performs poorly due to its patch‑plus‑stitching strategy, which makes the watermark visually obtrusive.

On ImageNet100, SSL‑Backdoor and CTRL achieve excellent fidelity (SSIM$>$0.99, LPIPS and DreamSim near zero). The patch becomes relatively small on high‑resolution $224\times224$ images, reducing visual impact for SSL‑Backdoor, while CTRL’s frequency‑domain perturbation maintains near‑perfect invisibility. Here, BLTO's fidelity degrades noticeably, likely because its optimization process is sensitive to image complexity. Moreover, NA remains inadequate, similar to its CIFAR10 behavior, confirming that its stitching‑based watermark is inherently unsuitable for fidelity.

\begin{table}[]
\centering
\caption{CL Model Accuracy Comparison on CIFAR10 and ImageNet100. \\ Encoder (Downstream) \%}
\resizebox{1.0\linewidth}{!}{
\begin{tabular}{c|c|cccc}
\hline
\textbf{Datasets}                     & \textbf{Methods}      & \textbf{SimCLR}       & \textbf{MoCo v2}       & \textbf{BYOL}          & \textbf{SimSiam}       \\ \hline
\multirow{5}{*}{\textbf{CIFAR10}}     & \textbf{Clean}        & \textbf{79.83 (79.8)} & \textbf{76.26 (77.38)} & \textbf{84.58 (83.21)} & \textbf{79.83 (80.49)} \\
                                      & \textbf{SSL-Backdoor} & 80.44 (81.3)          & 75.64 (77.84)          & 85.05 (83.71)          & 80.71 (80.57)          \\
                                      & \textbf{CTRL}         & 80.33 (80.8)          & 76.41 (77.54)          & 84.91 (82.99)          & 80.37 (78.93)          \\
                                      & \textbf{BLTO}         & 80.17 (80.09)         & 75.81 (77.81)          & 84.47 (82.84)          & 80.64 (79.34)          \\
                                      & \textbf{NA}           & 80.24 (80.86)         & 75.69 (77.43)          & 84.57 (83.48)          & 79.6 (80.04)           \\ \hline
\multirow{5}{*}{\textbf{ImageNet100}} & \textbf{Clean}        & \textbf{45.5 (49.32)} & \textbf{32.28 (37.24)} & \textbf{51.18 (55.44)} & \textbf{43.26 (48.62)} \\
                                      & \textbf{SSL-Backdoor} & 45.48 (49.18)         & 33.18 (36.7)           & 52.7 (54.04)           & 39.14 (40.46)          \\
                                      & \textbf{CTRL}         & 45.48 (48.54)         & 33.62 (36.6)           & 52.06 (55.34)          & 44.06 (49.92)          \\
                                      & \textbf{BLTO}         & 45.6 (48.68)          & 32.56 (38)             & 51.3 (55.04)           & 45.6 (49.26)           \\
                                      & \textbf{NA}           & 44.88 (47.78)         & 32.2 (37.16)           & 51.02 (54.18)          & 44.46 (49.16)          \\ \hline
\end{tabular}
}
\label{tab:utility}
\end{table}

\subsubsection{Model Utility} The training dataset is split into two parts: an unsupervised pretraining subset (80\%, unlabeled) and a downstream task fine-tuning subset (20\%, labeled). We first conduct 600 epochs of pretraining on four different CL models: SimCLR, MoCo v2, BYOL, and SimSiam. The encoder accuracy and downstream fine-tuning accuracy (in parentheses) are reported in \autoref{tab:utility}.

On CIFAR10, for all four watermarking methods, both encoder and downstream accuracies remain within a narrow margin (typically less than 1.5\%) of the clean baseline across all four models, indicating that none of the methods degrade model utility on this dataset.

On ImageNet100, most watermarking methods preserve model utility. CTRL, BLTO, and NA achieve encoder and downstream accuracies comparable to the clean baseline across all models, with differences generally within 2\%. However, SSL-Backdoor causes a notable performance drop on the SimSiam model: encoder accuracy falls from 43.26\% (clean) to 39.14\% (a 4.12\% decrease), and downstream accuracy drops from 48.62\% to 40.46\% (an 8.16\% decrease). This is likely because SimSiam lacks negative samples and momentum encoders, making it more sensitive to strong local perturbations introduced by the patch‑based SSL‑Backdoor watermark, especially on high‑resolution images. For SimCLR, MoCo v2, and BYOL, SSL-Backdoor shows no significant degradation.

\takeaway{SSL-Backdoor's fidelity is moderate on low‑resolution data but excellent on high‑resolution data. CTRL maintains near‑perfect invisibility across all resolutions. Furthermore, BLTO works well only on low‑resolution data and degrades noticeably on high‑resolution data. Notably, NA consistently fails in perceptual quality regardless of resolution. Regarding model utility, all four methods preserve encoder and downstream task performance, with the sole exception of SSL‑Backdoor on SimSiam with ImageNet100, which causes a noticeable drop.}

\subsection{Verifiability}
\label{Verifiability}
We use CIFAR10 and ImageNet100 as IP datasets, and introduce the similar but clean STL10 \cite{stl10} and ImageNette \cite{imagenette} datasets for comparison. \autoref{tab:sslbackdoor}–\autoref{tab:na} present the similarity differences $\Delta$ computed for SSL‑Backdoor, CTRL, BLTO and NA at the feature, soft label, and hard label levels. 
In each table, \textbf{bold} entries indicate $\Delta$ computed from models trained on the IP dataset, while non‑bold entries correspond to models trained on non‑IP datasets. Note that the soft label and hard label results are obtained under downstream tasks that share the same data domain as the pretraining task.

\begin{table}[htbp]
\centering
\caption{SSL-Backdoor's similarity differences $\Delta$ on IP dataset (CIFAR10 or ImageNet100) and non-IP datasets (others). \\ (Feature, Soft Label, Hard Label, $\times 10^{-2}$)}
\resizebox{1.0\linewidth}{!}{
\begin{tabular}{c|ccc|ccc|ccc|ccc}
\hline
\textbf{Models}      & \multicolumn{3}{c|}{\textbf{SimCLR}}           & \multicolumn{3}{c|}{\textbf{MoCo v2}}         & \multicolumn{3}{c|}{\textbf{BYOL}}             & \multicolumn{3}{c}{\textbf{SimSiam}}           \\ \hline
\textbf{CIFAR10}     & \textbf{15.72} & \textbf{6}    & \textbf{6.93} & \textbf{5.42} & \textbf{3.29} & \textbf{3.17} & \textbf{11.06} & \textbf{3.35} & \textbf{3.73} & \textbf{13.05} & \textbf{4.57} & \textbf{3.51} \\
\textbf{ImageNet100} & 4.17           & 0.69          & 0             & 2.24          & 5.65          & 0             & 2.95           & 1.82          & 0             & 2.97           & 3.26          & 0             \\
\textbf{STL10}       & 4.28           & -0.16         & 3.4           & 2.36          & 0.71          & -2.16         & 3.85           & 1.79          & 0.34          & 2.89           & 6.57          & 1.46          \\
\textbf{ImageNette}  & 3.33           & 2.44          & -10.17        & 2.33          & -5.31         & -10.62        & 4.08           & 1.23          & -7.81         & 5.52           & 3.48          & -7.28         \\ \hline
\textbf{CIFAR10}     & 0.25           & 0.33          & 0             & 0.05          & -0.41         & 0             & 0.22           & 0.22          & 0.19          & 0.15           & 0.3           & 0.18          \\
\textbf{ImageNet100} & \textbf{0.57}  & \textbf{0.24} & \textbf{0}    & \textbf{0.44} & \textbf{0.05} & \textbf{-0.2} & \textbf{0.49}  & \textbf{0.12} & \textbf{0}    & \textbf{0.39}  & \textbf{0.1}  & \textbf{-0.2} \\
\textbf{STL10}       & 0.16           & -0.56         & 0             & 0.01          & -0.54         & 0             & 0.2            & -0.03         & 0.4           & 0.3            & -0.37         & -0.2          \\
\textbf{ImageNette}  & 0.35           & 0.21          & 0             & 0.19          & 0.31          & 0             & 0.26           & 0.54          & -0.61         & 0.41           & 0.54          & 0.16          \\ \hline
\end{tabular}
}
\label{tab:sslbackdoor}
\end{table}

\subsubsection{SSL-Backdoor} On CIFAR10, SSL‑Backdoor demonstrates effective verification across all four model architectures. The IP $\Delta$ at the feature level are consistently larger than those of non‑IP datasets, and the margin remains clear at the soft label and hard label levels as well. This indicates that the patch‑based trigger, despite its simplicity, can be reliably learned by CL models when the image resolution is low. The watermark signal is strong enough to survive the unsupervised pretraining process and manifests as a clear density concentration in the output space.

On ImageNet100, however, SSL‑Backdoor fails entirely. For every model and every verification level, the IP $\Delta$ are near zero and cannot be distinguished from the non‑IP baselines, which also yield small or negative values. The underlying reason is the fixed size of the patch trigger. On CIFAR10's $32\times32$ images, the patch occupies a substantial area, creating a strong and learnable pattern. On ImageNet100's $224\times224$ images, the same patch becomes proportionally tiny, providing insufficient signal for the model to form a robust association with the target class. Consequently, no watermark effect is offered.

In summary, under settings with a fixed patch size, SSL‑Backdoor is a viable watermarking method for low‑resolution datasets. Its verifiability collapses when applied to high‑resolution data, suggesting that the patch size may need to be scaled appropriately for the target resolution.

\begin{table}[htbp]
\centering
\caption{CTRL's similarity differences $\Delta$ on IP dataset (CIFAR10 or ImageNet100) and non-IP datasets (others). \\ (Feature, Soft Label, Hard Label, $\times 10^{-2}$)}
\resizebox{1.0\linewidth}{!}{
\begin{tabular}{c|ccc|ccc|ccc|ccc}
\hline
\textbf{Models}      & \multicolumn{3}{c|}{\textbf{SimCLR}}             & \multicolumn{3}{c|}{\textbf{MoCo v2}}           & \multicolumn{3}{c|}{\textbf{BYOL}}               & \multicolumn{3}{c}{\textbf{SimSiam}}             \\ \hline
\textbf{CIFAR10}     & \textbf{31.49} & \textbf{30.87} & \textbf{12.99} & \textbf{0.74}  & \textbf{0.24}  & \textbf{0.18} & \textbf{12.01} & \textbf{25.5}  & \textbf{14.46} & \textbf{20.19} & \textbf{27.91} & \textbf{16.41} \\
\textbf{ImageNet100} & 0.13           & 0.07           & 0              & 0.36           & -1.23          & 0             & 0.06           & 0.09           & 0              & 0.001          & 0.32           & 0              \\
\textbf{STL10}       & 0.81           & 0.87           & 2.11           & 1.07           & 3.01           & 7.37          & 0.78           & 1.69           & 3.74           & 0.42           & 1.06           & 2.27           \\
\textbf{ImageNette}  & 0.12           & 0.04           & -0.55          & -0.005         & -0.01          & -0.18         & 0.08           & 0.12           & -0.32          & 0.14           & 0.13           & 0.16           \\ \hline
\textbf{CIFAR10}     & -0.12          & -0.71          & -2.48          & 0.07           & -0.12          & 0.59          & 0.53           & 0.29           & 1.16           & -0.21          & 0.004          & -1.16          \\
\textbf{ImageNet100} & \textbf{0.01}  & \textbf{-0.02} & \textbf{0}     & \textbf{0.008} & \textbf{0.004} & \textbf{0}    & \textbf{0.02}  & \textbf{0.006} & \textbf{0}     & \textbf{0.01}  & \textbf{0.005} & \textbf{0}     \\
\textbf{STL10}       & -0.004         & -0.03          & 0              & -0.07          & -0.29          & -0.2          & -0.03          & -0.08          & 0              & 0.03           & -0.007         & 0              \\
\textbf{ImageNette}  & -0.009         & 0.009          & 0              & -0.01          & -0.004         & 0.16          & -0.02          & -0.005         & 0.31           & -0.02          & -0.01          & 0              \\ \hline
\end{tabular}
}
\label{tab:ctrl}
\end{table}

\subsubsection{CTRL} On CIFAR10, CTRL shows mixed but generally positive verification performance. For SimCLR, BYOL, and SimSiam, the IP $\Delta$ at all three verification levels is substantially larger than that of non‑IP datasets, indicating successful watermark embedding. However, CTRL completely fails on MoCo v2: the IP $\Delta$ are even smaller than some non‑IP $\Delta$, suggesting that MoCo v2's momentum CL mechanism may be less sensitive to the static frequency‑domain perturbations used by CTRL. This architecture dependency is a notable limitation.

On ImageNet100, CTRL fails across all models and verification levels. 
The reason lies in static frequency‑domain perturbations. On low‑resolution CIFAR10, the high‑frequency noise is relatively concentrated and can be easily captured by the model during pretraining, leading to a measurable watermark effect. On high‑resolution ImageNet100, the same fixed perturbation becomes sparsely distributed across a much larger pixel space, making it difficult for the model to learn and retain the pattern. As a result, the watermark signal is lost.

In summary, CTRL offers verifiable watermarks only under favorable conditions (low resolution, compatible architectures like SimCLR/BYOL/SimSiam). Its sensitivity to both model architecture and dataset resolution makes it less reliable for general‑purpose dataset watermarking.

\begin{table}[htbp]
\centering
\caption{BLTO's similarity differences $\Delta$ on IP dataset (CIFAR10 or ImageNet100) and non-IP datasets (others). \\ (Feature, Soft Label, Hard Label, $\times 10^{-2}$)}
\resizebox{1.0\linewidth}{!}{
\begin{tabular}{c|ccc|ccc|ccc|ccc}
\hline
\textbf{Models}      & \multicolumn{3}{c|}{\textbf{SimCLR}}              & \multicolumn{3}{c|}{\textbf{MoCo v2}}            & \multicolumn{3}{c|}{\textbf{BYOL}}               & \multicolumn{3}{c}{\textbf{SimSiam}}             \\ \hline
\textbf{CIFAR10}     & \textbf{12.54} & \textbf{17.11} & \textbf{14.98}  & \textbf{7.4}   & \textbf{17.05} & \textbf{11.44} & \textbf{13.19} & \textbf{20.35} & \textbf{16.66} & \textbf{12.19} & \textbf{14.02} & \textbf{16.21} \\
\textbf{ImageNet100} & 2.02           & 0.35           & 0               & 2.15           & 3.29           & 0              & 2.36           & 1.01           & 0              & 1.65           & 2.57           & 0              \\
\textbf{STL10}       & 4.3            & -1.55          & 6.32            & 2.77           & 1.53           & 6.59           & 4.37           & 2.73           & 10.43          & 1.1            & 3.78           & 11.07          \\
\textbf{ImageNette}  & 1.52           & 1.78           & -0.18           & 0.78           & 0.22           & 0.53           & 1.31           & 1.47           & 0.48           & 2.32           & 2.18           & 2.25           \\ \hline
\textbf{CIFAR10}     & 11.9           & 5.23           & -11.51          & 3.22           & -10.02         & -0.39          & 10.21          & 7.87           & -11.03         & 8.49           & 1.4            & 10.38          \\
\textbf{ImageNet100} & \textbf{21.65} & \textbf{10.75} & \textbf{-20.18} & \textbf{10.96} & \textbf{6.28}  & \textbf{1.59}  & \textbf{22.7}  & \textbf{10.43} & \textbf{0.2}   & \textbf{23.85} & \textbf{16.63} & \textbf{-0.99} \\
\textbf{STL10}       & 5.75           & 2.54           & 0.6             & 1.23           & -7.68          & -3.55          & 10.67          & 6.29           & 1.79           & 9.86           & 5.63           & 2.78           \\
\textbf{ImageNette}  & 4.5            & 2.89           & -2.21           & 2.57           & 1.59           & -2.77          & 3.45           & 3.21           & 2.16           & 6.02           & 4.68           & 1.32           \\ \hline
\end{tabular}
}
\label{tab:blto}
\end{table}

\subsubsection{BLTO} On CIFAR10, BLTO achieves strong and consistent verification across all four model architectures and all three verification levels. The IP $\Delta$ is uniformly larger than non‑IP baselines, with clear margins. 
The adaptive nature of BLTO, where the trigger is tailored to the target class and dataset through a bi-level optimization, ensures that the watermark signal is both learnable and discriminative.

On ImageNet100, BLTO remains effective at the feature and soft label levels for all models, with IP $\Delta$ substantially exceeding non‑IP $\Delta$. However, at the hard label level, the signal becomes less reliable for some architectures (e.g., SimCLR, BYOL, SimSiam), where the $\Delta$ are sometimes lower than or close to non‑IP $\Delta$. This indicates that while the watermark creates a strong density concentration in the continuous output spaces (feature vectors and softmax probability vectors), the discrete hard labels may not fully capture the fine‑grained clustering effect. Nevertheless, given that most verification scenarios allow access to soft labels or features, BLTO remains highly effective.

In summary, BLTO provides the most balanced verifiability across datasets and different output levels. Its performance on high‑resolution data is particularly noteworthy, as it maintains a clear watermark signal where other methods fail.

\begin{table}[htbp]
\centering
\caption{NA's similarity differences $\Delta$ on IP dataset (CIFAR10 or ImageNet100) and non-IP datasets (others). \\ (Feature, Soft Label, Hard Label, $\times 10^{-2}$)}
\resizebox{0.95\linewidth}{!}{
\begin{tabular}{c|ccc|ccc|ccc|ccc}
\hline
\textbf{Models}      & \multicolumn{3}{c|}{\textbf{SimCLR}}            & \multicolumn{3}{c|}{\textbf{MoCo v2}}             & \multicolumn{3}{c|}{\textbf{BYOL}}              & \multicolumn{3}{c}{\textbf{SimSiam}}            \\ \hline
\textbf{CIFAR10}     & \textbf{54.51} & \textbf{80.28} & \textbf{24.1} & \textbf{13.56} & \textbf{35.41} & \textbf{-21.09} & \textbf{55.81} & \textbf{83.86} & \textbf{24.3} & \textbf{37.8} & \textbf{55.75} & \textbf{-1.58} \\
\textbf{ImageNet100} & 4.41           & 5.11           & 0             & 1.99           & 9.13           & 0               & 10.32          & 5.6            & 0             & 5.15          & 1.48           & 0              \\
\textbf{STL10}       & 5.1            & 9.76           & -19.72        & 4.09           & 1.47           & -5.79           & 5.11           & 0.87           & -14.15        & 1.1           & -3             & -15.39         \\
\textbf{ImageNette}  & 2.27           & 2.47           & 10.06         & -2.03          & -354.3         & 8.56            & -0.52          & -0.22          & 4.51          & -3.02         & -1.37          & 13.06          \\ \hline
\textbf{CIFAR10}     & 1.16           & 2.56           & 0.84          & 0.59           & -1.44          & 0.2             & 1.01           & 2.09           & 0.58          & 0.23          & 1.07           & -0.37          \\
\textbf{ImageNet100} & \textbf{28.29} & \textbf{35.89} & \textbf{0.4}  & \textbf{14.36} & \textbf{19.29} & \textbf{0.99}   & \textbf{51.43} & \textbf{65.77} & \textbf{0.4}  & \textbf{59.9} & \textbf{84.62} & \textbf{0.8}   \\
\textbf{STL10}       & 4.69           & 6.34           & -1.19         & 2.04           & 8.31           & -0.79           & 2.55           & 1.37           & -1.38         & 2.95          & 2.52           & -2.55          \\
\textbf{ImageNette}  & 4.28           & 1.38           & 0             & 1.62           & 1.01           & -4.21           & 3.43           & 1.113          & -11.45        & 2.95          & 1.23           & 1.99           \\ \hline
\end{tabular}
}
\label{tab:na}
\end{table}

\subsubsection{NA} On CIFAR10, NA produces exceptionally large IP $\Delta$ at the feature and soft label levels across all architectures, far exceeding those of non‑IP datasets. The margins are substantially larger than those achieved by any other method, indicating an extremely strong watermark signal. At the hard label level, the results are less consistent: for some architectures (e.g., MoCo v2, SimSiam), the IP $\Delta$ is negative or close to zero, suggesting that the discrete label output may not capture the watermark's effect. Nonetheless, the feature and soft label signals are so pronounced that verification is unambiguous.

On ImageNet100, NA performs even better. The IP $\Delta$ at feature and soft label levels reach remarkably high magnitudes across all models, with clear separation from non‑IP datasets. The hard label level yields only small positive values, but the feature and soft label signals are sufficiently discriminative to enable reliable verification. This demonstrates that NA's patch‑plus‑stitching watermark, despite its poor perceptual quality, creates an extremely robust and scalable watermark that becomes even more effective on high‑resolution data. The stitching operation introduces a global structural change that is consistently learned by contrastive models regardless of resolution.

In summary, NA achieves the highest verifiability among all four methods, delivering strong watermark signals on both CIFAR10 and ImageNet100. The trade‑off, as shown in the fidelity evaluation, is a significant degradation in perceptual quality, making NA suitable only for applications where visual imperceptibility is not a primary concern.


\begin{figure*}[htbp]
    \centering
    \includegraphics[width=0.99\linewidth]{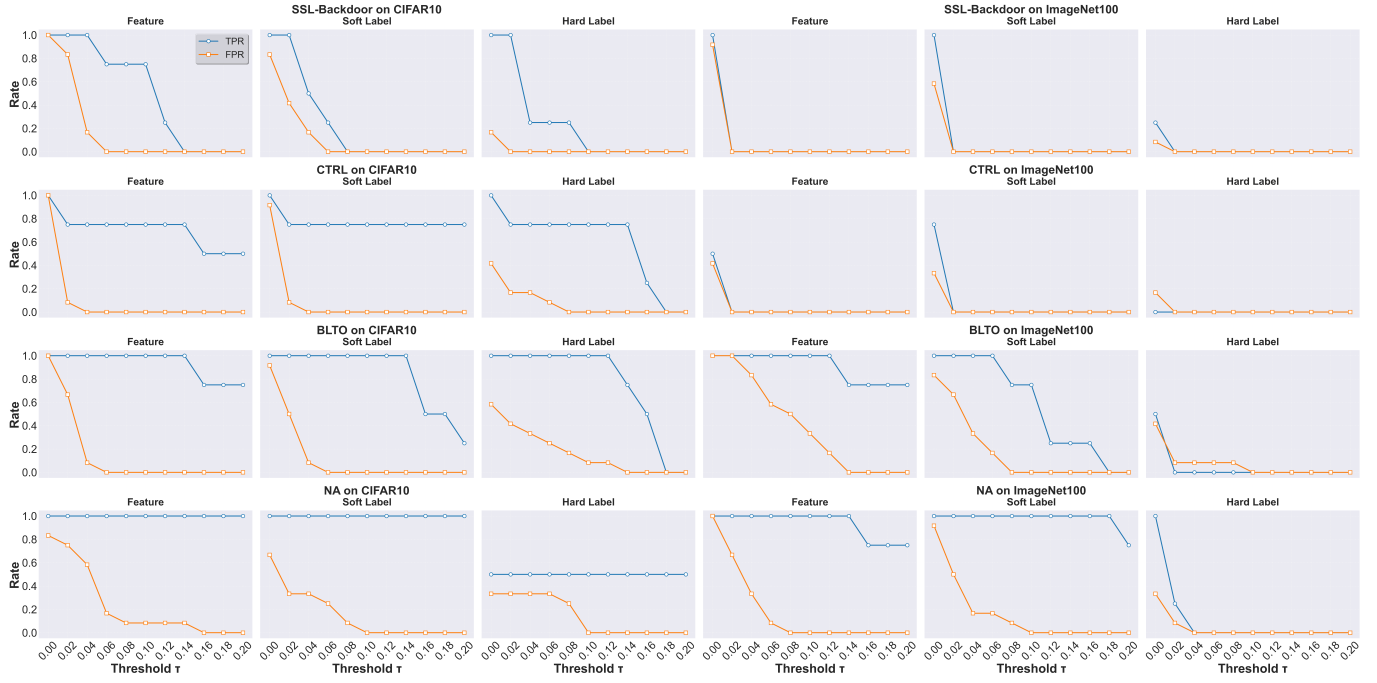} 
    \caption{Comparison of TPR and FPR for SSL-Backdoor, CTRL, BLTO and NA under different thresholds. (Feature / Soft Label / Hard Label levels)}
    \label{fig:threshold}
\end{figure*}

\subsubsection{Hypothesis Testing}
Previously, we mainly used $\Delta$ to demonstrate the distinguishability from IP and non-IP datasets through each of the methods. However, it is not a standard metric when evaluating the True Positive Rate (TPR) and False Positive Rate (FPR).

Hypothesis testing provides a rigorous framework to compute the TPR and FPR under a given decision threshold. However, the optimal threshold may vary depending on the watermark method, verification level (feature, soft label, hard label), dataset, and model architecture. In practice, a dataset owner knows the dataset they own and can infer the verification level from the suspect model's API outputs (e.g., feature vectors, soft labels, or hard labels). They typically do not know the exact CL architecture (e.g., SimCLR, MoCo v2) used to train the suspect model. Therefore, we aggregate results across all four model architectures for each method, dataset, and verification level, reflecting a realistic black-box scenario where the architecture is unknown.

Following the hypothesis testing framework defined in \autoref{Verifiability}, we compute for each threshold $\tau$ the TPR and FPR based on the 16 suspect models (4 IP-trained positives, 12 non-IP-trained negatives) per setting. \autoref{fig:threshold} presents the TPR/FPR curves.

On CIFAR10, SSL‑Backdoor achieves moderate verification performance: its TPR is reasonably high at small thresholds but declines quickly as the threshold increases, with FPR dropping to zero at a relatively low threshold. CTRL exhibits stable but imperfect verification, maintaining moderate TPR across a wide threshold range at feature and soft label levels, while hard label TPR degrades at higher thresholds; its FPR becomes zero at a small threshold. BLTO demonstrates excellent verification: at feature and soft label levels, TPR remains perfect for thresholds up to around 0.14, and hard label TPR is also high for thresholds up to 0.12 before gradually decreasing. NA achieves the strongest performance, maintaining perfect TPR at feature and soft label levels across nearly all thresholds up to 0.2, with FPR approaching zero around 0.1; its hard label TPR is consistently limited. Overall, on CIFAR10, a threshold of 0.10–0.14 works well for BLTO and NA, while SSL‑Backdoor and CTRL require smaller thresholds (e.g., 0.04–0.06) to balance TPR and FPR.

On ImageNet100, the performance gap becomes pronounced. SSL‑Backdoor and CTRL are almost completely ineffective: their TPR drops to zero for any positive threshold, confirming their resolution sensitivity observed earlier. BLTO remains effective at the feature level, maintaining perfect TPR for thresholds up to about 0.12, and moderately effective at the soft label level for thresholds up to 0.08, but hard label verification completely fails. Note that NA continues to perform strongly, preserving perfect TPR at feature and soft label levels for thresholds up to 0.14–0.18, with FPR reaching zero around 0.08–0.10; hard label verification only works at the smallest threshold. For ImageNet100, we recommend a threshold of 0.10–0.12 for BLTO feature level verification and 0.10–0.14 for NA feature/soft label verification.

\takeaway{Verifiability varies by method. SSL‑Backdoor and CTRL only work on low‑resolution data (CTRL also fails on MoCo v2). Whereas, BLTO and NA are effective across datasets at feature/soft label levels, with NA yielding stronger signals but poorer fidelity. Threshold selection matters: on CIFAR10, use 0.04–0.14 (smaller for SSL‑Backdoor/CTRL); on ImageNet100, only BLTO/NA work (threshold 0.08–0.12). Choose NA for max verifiability or BLTO for better fidelity.}

\begin{table}[htbp]
\centering
\caption{Four Methods' similarity differences $\Delta$ on IP dataset (CIFAR10) for different downstream tasks. \\ (Soft Label, Hard Label, $\times 10^{-2}$)}
\resizebox{0.95\linewidth}{!}{
\begin{tabular}{c|c|cc|cc|cc|cc}
\hline
\textbf{Methods}                       & \textbf{Models}     & \multicolumn{2}{c|}{\textbf{SimCLR}} & \multicolumn{2}{c|}{\textbf{MoCo v2}} & \multicolumn{2}{c|}{\textbf{BYOL}} & \multicolumn{2}{c}{\textbf{SimSiam}} \\ \hline
\multirow{3}{*}{\textbf{SSL-Backdoor}} & \textbf{CIFAR10}    & 6      & 6.93      & 3.29     & 3.17     & 3.35    & 3.73   & 4.57     & 3.51    \\
                                       & \textbf{STL10}      & 9.59            & -1.92              & 2.37              & -0.53             & 4.08             & -10.68          & 4.74              & 0.36             \\
                                       & \textbf{ImageNette} & 20.1            & 9.06               & 2.15              & -6.28             & 14.69            & -1.53           & 13.82             & -3.42            \\ \hline
\multirow{3}{*}{\textbf{CTRL}}         & \textbf{CIFAR10}    & 30.87           & 12.99              & 0.24              & 0.18              & 25.5             & 14.46           & 27.91             & 16.41            \\
                                       & \textbf{STL10}      & 10.46           & 12.74              & 0.72              & 2.99              & 8.48             & 11.49           & 10.85             & 14.91            \\
                                       & \textbf{ImageNette} & 11.86           & 14.95              & 0.31              & -2.05             & 8.39             & 6.2             & 3.19              & 2.82             \\ \hline
\multirow{3}{*}{\textbf{BLTO}}         & \textbf{CIFAR10}    & 17.11           & 14.98              & 17.05             & 11.44             & 20.35            & 16.66           & 14.02             & 16.21            \\
                                       & \textbf{STL10}      & 30.4            & 18.89              & 30.63             & 19.3              & 38.38            & 21.24           & 16.5              & 19.58            \\
                                       & \textbf{ImageNette} & 32.79           & 16.68              & 28.66             & 20.26             & 42.95            & 17.77           & 25.92             & 12.64            \\ \hline
\multirow{3}{*}{\textbf{NA}}           & \textbf{CIFAR10}    & 80.28           & 24.1               & 35.41             & -21.09            & 83.86            & 24.3            & 55.75             & -1.58            \\
                                       & \textbf{STL10}      & 83.44           & 19.37              & 27.77             & -28.5             & 70.84            & 19.35           & 57.96             & 1.43             \\
                                       & \textbf{ImageNette} & 42.44           & 31.79              & 33.16             & 21.16             & 67.89            & 12.02           & 42.21             & 22.04            \\ \hline
\end{tabular}
}
\label{tab:robustness}
\end{table}

\subsection{Robustness}
Pretrained CL models commonly serve as foundation models for various downstream tasks. To evaluate whether watermarks survive after fine-tuning to different tasks, we pretrain models on the watermarked CIFAR10 dataset (where all four methods are verifiable), then fine-tune them on two distinct downstream datasets: STL10 and ImageNette, using only 20\% labeled data without watermarks. The similarity differences $\Delta$ (soft and hard label levels) for the IP dataset (CIFAR10) after fine-tuning are reported in \autoref{tab:robustness}. The similarity differences $\Delta$ that are comparable to or higher than the original CIFAR10 baseline (i.e., fine-tuning on the same CIFAR10 task) indicate good robustness.

\subsubsection{SSL-Backdoor} For SSL-Backdoor, soft label $\Delta$ on the two different downstream tasks is generally comparable to or slightly higher than the CIFAR10 baseline, suggesting that its patch-based watermark retains some transferability across natural data domains. However, hard label $\Delta$ often becomes much lower or even negative, indicating that discrete predictions are less reliable under distribution shift. This asymmetry arises because SSL-Backdoor’s trigger creates a strong semantic association that survives in continuous soft label space, but the hard label decision boundary is more sensitive to domain changes.

\subsubsection{CTRL} It exhibits poor robustness. For most architectures, soft label and hard label $\Delta$ on STL10 and ImageNette drop substantially compared to the CIFAR10 baseline. This fragility is inherent to CTRL’s frequency-domain perturbations: they are semantically unrelated to the target class and act as static noise. When the model adapts to a new data distribution, it quickly forgets these meaningless correlations, causing the watermark signal to degrade.

\subsubsection{BLTO} It demonstrates excellent robustness. For all architectures, both soft label and hard label $\Delta$ on STL10 and ImageNette are consistently higher than those on the original CIFAR10 task. This counterintuitive improvement stems from BLTO’s adaptive optimization mechanism. BLTO generates watermarks via a bi‑level optimization that tightly couples the trigger with the target class’s intrinsic features. During fine‑tuning on a new downstream task, the model gradually forgets task‑specific features of the original pretraining data but selectively retains the strongly regularized watermark features, because they are more discriminative and less entangled with the original data distribution. As a result, watermark samples become even more separated from other classes in the new feature space, leading to an increased $\Delta$.

\subsubsection{NA} It shows strong robustness, particularly at the soft label level. On both downstream tasks, soft label $\Delta$ remains comparable to the high CIFAR10 baseline, and in some cases even slightly higher. Hard-label $\Delta$ varies across architectures and tasks, and can become positive on some downstream datasets (e.g., ImageNette) even if it was negative on the original CIFAR10 baseline. Note that NA's patch-plus-stitching watermark creates a very strong and perceptually salient signal. This external pattern is highly discriminative, allowing the model to retain it during fine-tuning. However, because the watermark is not adaptively integrated with the target class’s intrinsic representations, it does not become further enhanced. The model simply preserves the existing strong signal, resulting in $\Delta$ values that are generally comparable to the baseline.

\takeaway{Robustness after fine-tuning varies across methods. SSL‑Backdoor is moderately robust only at the soft label level; CTRL is fragile and loses effectiveness. BLTO shows excellent robustness with $\Delta$ consistently higher than baseline, due to its adaptive feature coupling. Also, NA shows strong robustness, maintaining high soft label $\Delta$, though hard label signals are more variable. For watermark persistence under model adaptation, BLTO and NA are preferred.}

\section{Conclusion}
This work first systematically evaluates six data-poisoning-only backdoor attacks under a unified CL framework, identifying that PoisonedEncoder and CorruptEncoder are fundamentally unsuitable for encoder-level backdoor injection due to their dependence on downstream task class information, while SSL-Backdoor, CTRL, BLTO, and NA exhibit varying degrees of downstream task‑independent backdoor effectiveness. We then propose to repurpose these four methods as dataset watermarking and design a multi‑level verification scheme that enables ownership verification at feature, soft label, and hard label levels. Through extensive experiments on four CL models across different datasets, we assess their fidelity, verifiability, and robustness. Our results reveal clear trade‑offs: BLTO and NA deliver the best overall watermarking performance, with BLTO offering balanced fidelity and verifiability, and NA achieving the strongest verifiability at the cost of poor perceptual quality. SSL‑Backdoor and CTRL are effective only under limited conditions (e.g., low‑resolution data or specific architectures). To the best of our knowledge, this is the first work to systematically benchmark backdoor attacks for dataset watermarking in CL, providing practical guidance for dataset IP protection.


%





\ifCLASSOPTIONcaptionsoff
  \newpage
\fi



%



\bibliographystyle{IEEEtran}
\bibliography{refs}

\vspace{-1.5cm}
\begin{IEEEbiography}[{\includegraphics[width=1in,height=1.25in,clip,keepaspectratio]{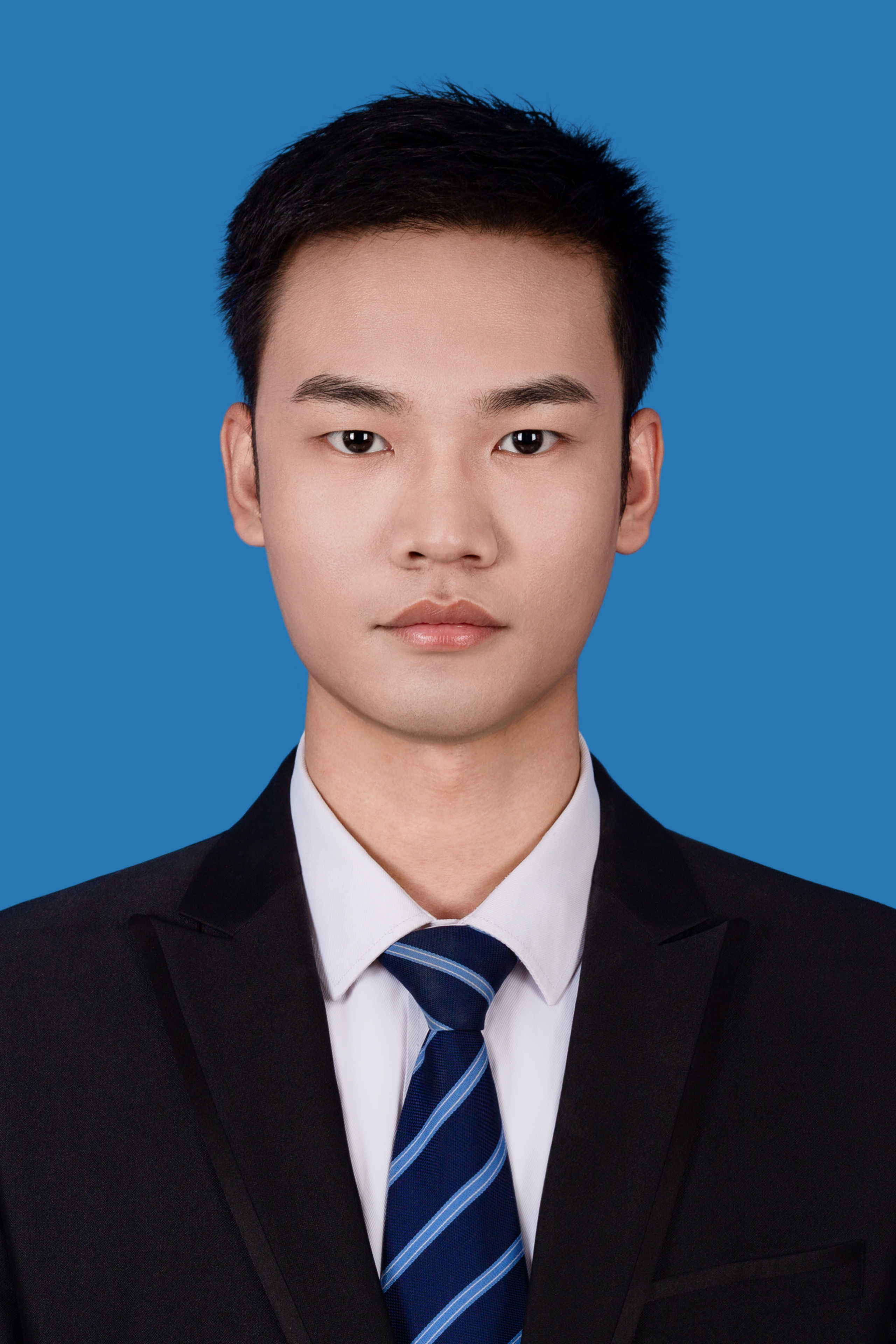}}]{Zhiyang~Dai} received the bachelor's degree in Qian Xuesen College from Nanjing University of Science and Technology, Nanjing, China, in 2021, where he is currently pursuing the Ph.D. degree with the School of Cyber Science and Engineering from Nanjing University of Science and Technology, Nanjing, China. His research interests include AI privacy and security.
\end{IEEEbiography}

\vspace{-1.2cm}
\begin{IEEEbiography}[{\includegraphics[width=1in,height=1.25in,clip,keepaspectratio]{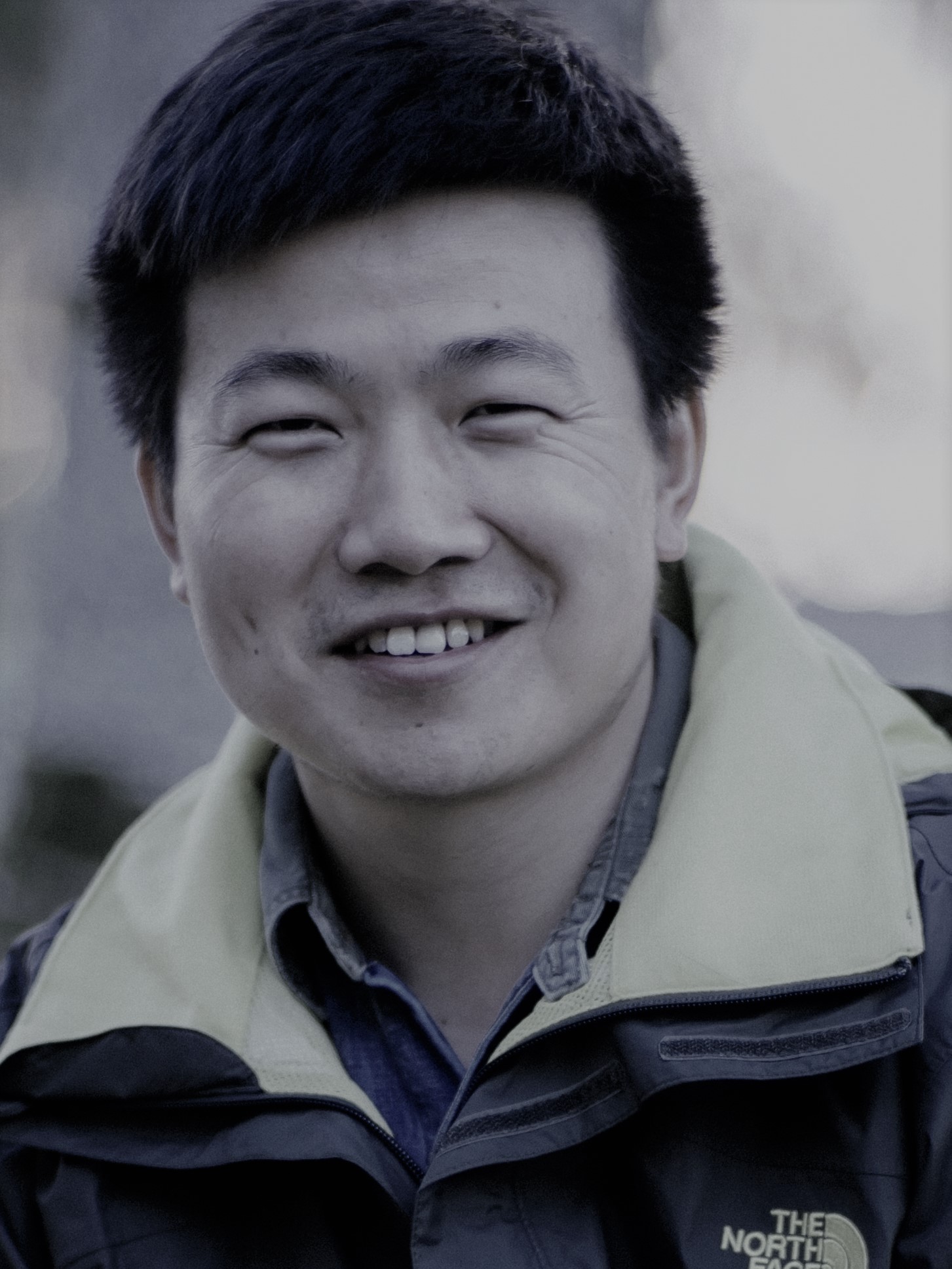}}]{Yansong~Gao} (Senior Member, IEEE) is a Lecturer at the University of Western Australia. He received his M.Sc degree from the University of Electronic Science and Technology of China and a Ph.D. degree from the University of Adelaide, Australia. His current research interests are AI security and privacy, hardware security, and system security. He serves as an Associate Editor of \sc{IEEE Transactions on Information Forensics and Security, IEEE Transactions on Neural Networks and Learning Systems}.
\end{IEEEbiography}

\vspace{-1.2cm}
\begin{IEEEbiography}[{\includegraphics[width=1in,height=1.25in,clip,keepaspectratio]{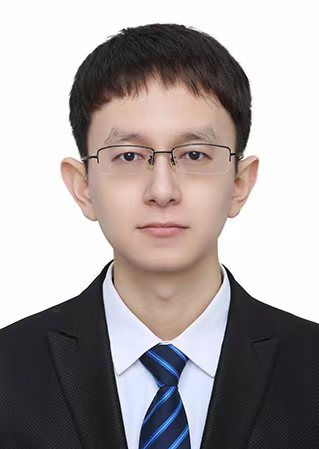}}]{Boyu~Kuang} received the Ph.D. degree in Cyber Space Security from Nanjing University of Science and Technology in 2023. He is currently an Associate Professor of Nanjing University of Science and Technology, China. His research interests include IoT security and vulnerability mining.
\end{IEEEbiography}

\vspace{-1.2cm}
\begin{IEEEbiography}[{\includegraphics[width=1in,height=1.25in,clip,keepaspectratio]{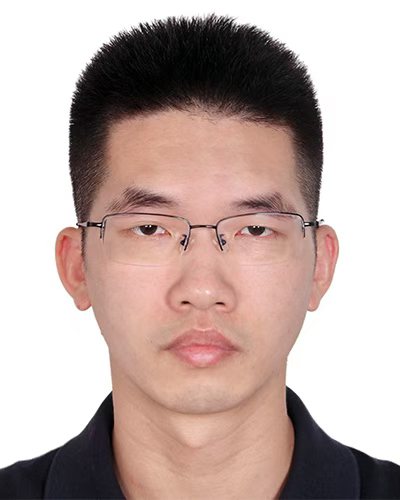}}]{Haodong~Li} received the B.S. and Ph.D. degrees from Sun Yat-sen University, Guangzhou, China, in 2012 and 2017, respectively. He is currently an Associate Professor with the College of Electronics and Information Engineering, Shenzhen University, Shenzhen, China. His current research interests include multimedia forensics, image processing, and machine learning.
\end{IEEEbiography}

\vspace{-1.2cm}
\begin{IEEEbiography}[{\includegraphics[width=1in,height=1.25in,clip,keepaspectratio]{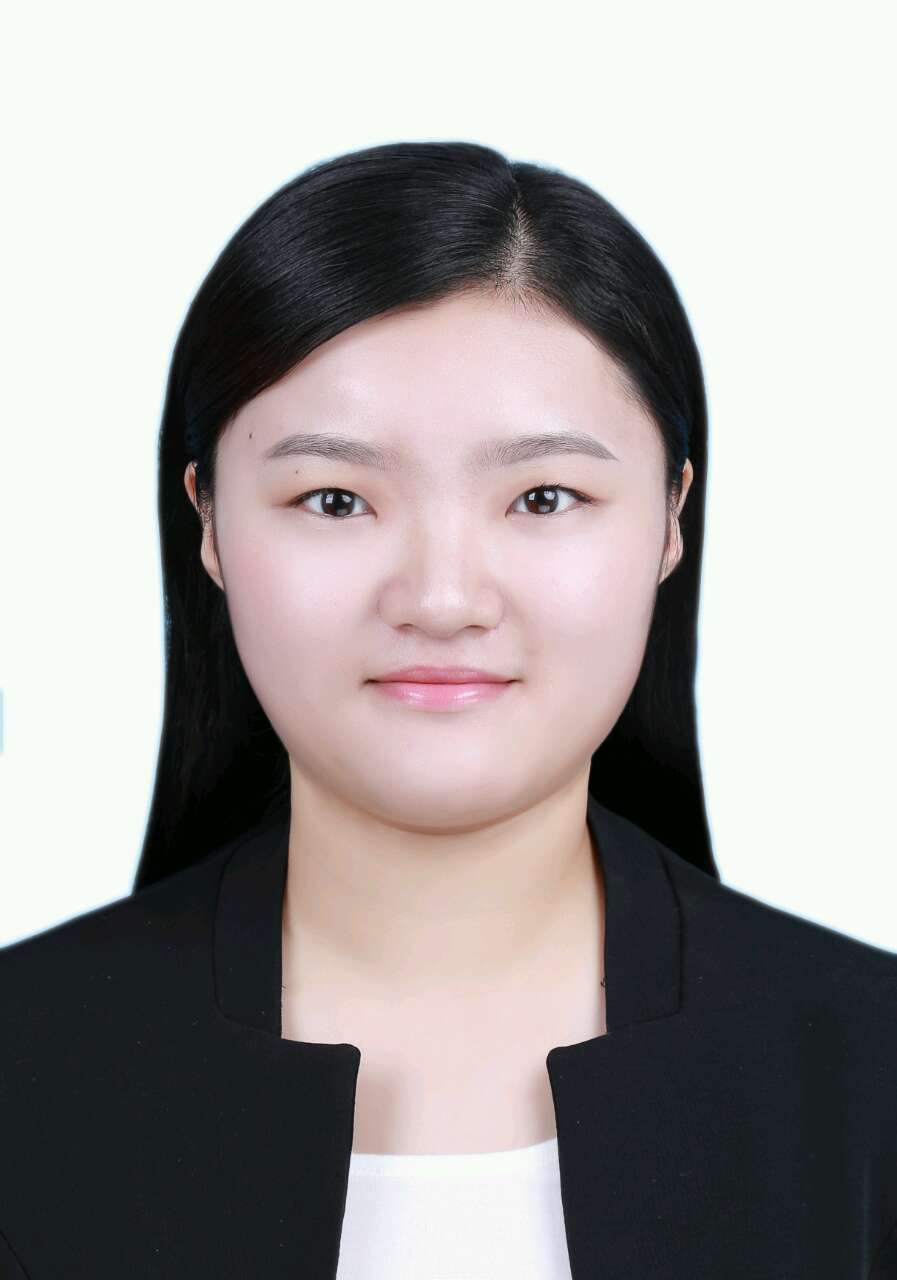}}]{Qi~Chang} received the B.S. and Ph.D. degrees from Beijing Jiaotong University, Beijing, China, in 2016 and 2022, respectively. She was a visiting Ph.D. student with the National Institute of Informatics (NII), Tokyo, Japan, in 2018. She is currently an Assistant Professor with the College of Electronics and Information Engineering, Shenzhen University, Shenzhen, China. Her research interests are image processing and information hiding.
\end{IEEEbiography}

\vspace{-1.2cm}
\begin{IEEEbiography}[{\includegraphics[width=1in,height=1.25in,clip,keepaspectratio]{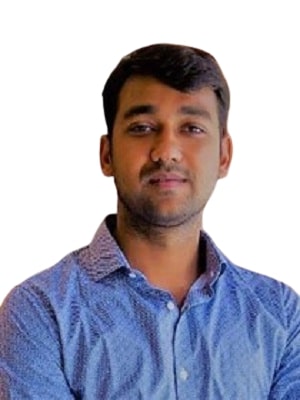}}]{Gaurav~Varshney} received his M.Tech and Ph.D. from IIT Roorkee. His current research interests are anti-phishing, internet security (protocols and policies), email security, web security, vulnerability assessment, threat modeling, network security, and cyber frauds. He has worked as a Research Intern at TRDDC Pune, an Engineer at Qualcomm, a Visiting Scholar at SUNY Albany (NY, USA), a Visiting Fellow at SUTD Singapore, and a Visiting Staff at the University of Newcastle, Australia.
\end{IEEEbiography}

\vspace{-1.2cm}
\begin{IEEEbiography}[{\includegraphics[width=1in,height=1.25in,clip,keepaspectratio]{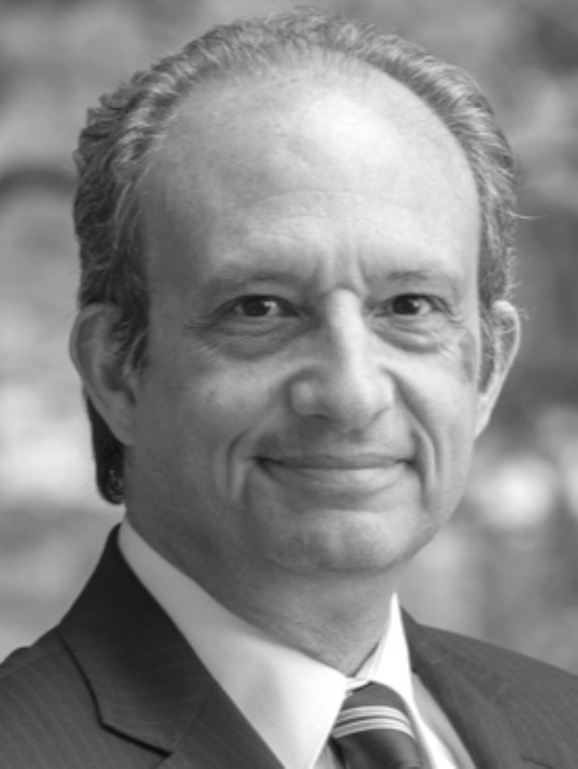}}] {Derek Abbott}(M'85–-SM'99–-F'05) was born in South Kensington, London, U.K. He received the B.Sc.~(Hons.) degree in physics from Loughborough University, U.K., in 1982, and the Ph.D.~degree in electrical and electronic engineering from the University of Adelaide, Australia, in 1997, under K.~Eshraghian and B.~R.~Davis. His research interests include the areas of multidisciplinary physics and electronic engineering applied to complex systems. His research programs span a number of areas including security, stochastics, game theory, security, photonics, energy policy, biomedical engineering, and computational neuroscience. He is a fellow of the Institute of Physics, U.K., an Honorary Fellow of Engineers Australia and Australian Laureate Fellow. He received a number of awards, including the South Australian Tall Poppy Award for Science, in 2004, an Australian Research Council Future Fellowship, in 2012, the David Dewhurst Medal, in 2015 the Barry Inglis Medal, in 2018, and the M.~A.~Sargent Medal for eminence in engineering, in 2019. He has served as an Editor and/or a Guest Editor for a number of journals, including the {\sc IEEE Journal of Solid-State Circuits}, {\it Journal of Optics B}, {\it Chaos}, {\it Fluctuation and Noise Letters,} {\it Royal Society OS}, {\sc Proceedings of the IEEE}, and the {\sc IEEE Photonics Journal}.
\end{IEEEbiography}

\vspace{-1.2cm}
\begin{IEEEbiography}[{\includegraphics[width=1in,height=1.25in,clip,keepaspectratio]{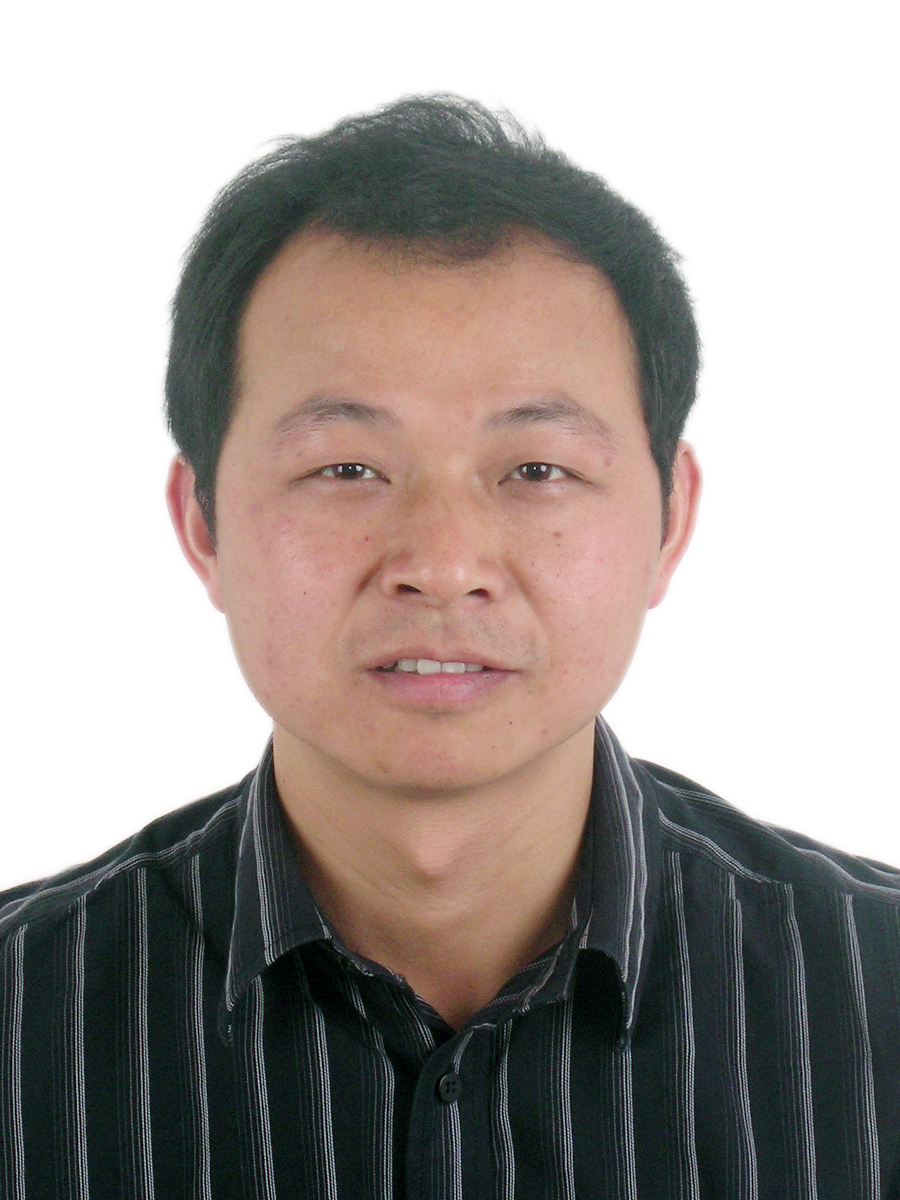}}]{Anmin Fu} (Member, IEEE) received the Ph.D. degree in Information Security from Xidian University, Xi’an, China, in 2011. He is currently a Professor of Nanjing University of Science and Technology, China. Dr Fu’s research interests include IoT Security, Cloud Computing Security, and Privacy Preserving. He has published more than 100 technical papers, including international journals and conferences, such as IEEE TDSC, IEEE TIFS, IEEE S\&P, ACM CCS, USENIX Security, and NDSS.
\end{IEEEbiography}

\end{document}